\documentclass[prd,showpacs,preprintnumbers,nofootinbib]{revtex4-1}
\usepackage{amssymb}
\usepackage{amsmath}
\usepackage{subeqnarray}
\usepackage[T1]{fontenc}
\usepackage[french]{babel}

\newcommand{\beq} {\begin{equation}}
\newcommand{\eeq} {\end{equation}}
\newcommand{\bea} {\begin{eqnarray}}
\newcommand{\eea} {\end{eqnarray}}
\newcommand{\bsea} {\begin{subeqnarray}}
\newcommand{\esea} {\end{subeqnarray}}
\newcommand{\nn}  {\nonumber}
\newcommand{\ga}{\gamma}
\newcommand{\ba}{\beta}
\newcommand{\da}{\delta}
\newcommand{\al}{\alpha}
\newcommand{\la}{\lambda}
\newcommand{\eps}{\epsilon}
\newcommand{\ka}{\kappa}
\newcommand{\sig}{\sigma}
\newcommand{\U}{\hat U}
\renewcommand{\l}[1]{\left#1}
\renewcommand{\r}[1]{\right#1}

\newcommand{\ud}{\mathrm{d}}
\newcommand{\si}{ \mathrm{si}}

\newcommand{\half}{\frac{1}{2}}

\begin{document}

\title{Einstein-Maxwell-Dilaton theories with a Liouville potential}
\author{Christos Charmousis$^{1,2)}$, Blaise Gout\'eraux$^{1)}$ and Jiro Soda$^{3)}$}
\affiliation{1) LPT Orsay, Univ. Paris-Sud, CNRS : UMR 8627, B\^at. 210, 91405 Orsay, CEDEX, France}
\affiliation{ 2) Laboratoire de Math\'ematiques et Physique Th\'eorique (LMPT) CNRS : UMR 6083\\
 Universit\'e Fran\c cois Rabelais - Tours}
\affiliation{3)Yukawa Institute for Theoretical Physics, Department of Physics,  Kyoto University, Kyoto, 606-8501, Japan}
\pacs{04.20.Jb}
\preprint{arXiv:gr-qc/0905.3337v3}

\begin{abstract}
We find and analyse solutions of Einstein's equations in arbitrary dimensions and in the presence of a scalar field with a Liouville potential coupled to a Maxwell field. We consider spacetimes of cylindrical symmetry or again  subspaces of dimension $d-2$ with constant curvature and analyse in detail the field equations and manifest their symmetries.
The field equations of the full system are shown to reduce to a single or  couple of ODE's which can be used to solve analytically or numerically the theory for the symmetry at hand. Further solutions can also be generated by a solution generating technique akin to the EM duality in the absence of a cosmological constant. We then find and analyse explicit solutions including black holes and gravitating solitons  for the case of four dimensional relativity and the higher-dimensional oxydised 5-dimensional spacetime. The general solution is obtained for a certain relation between couplings in the case of cylindrical symmetry.
\end{abstract}

\maketitle

\section{Introduction}

Black hole solutions capture the essence of general relativity in that they are relatively simple but not trivial, vacuum solutions of Einstein's field equations with multiple applications in different domains of gravitational physics. Black holes appeared, at least as mathematical solutions of Einstein equations, very early on in the development of the theory\cite{scwh}. Initially however, it was hoped by Einstein and disciples that classical, gravitating, and everywhere regular particle solutions, in other words solitons, could be found in the context of general relativity theory. For a long time there was a general consensus that black holes were just a bizarre and singular mathematical artefact of general relativity and not physical solutions. In a way they were seen as a contradiction to the Machian principle which inspired  Einstein in his formulation of General Relativity. Two main mathematical results tilted scientific bias in favour for black holes.  First, the soliton paradigm did not work for a non-asymptotically flat spacetime, in that if a solution of Einstein's equations is asymptotically flat, topologically trivial and globally stationary, then it is just flat spacetime \cite{Einstein} (for a nice and precise introduction see \cite{maison}). Secondly, and this radically changed the black hole concensus, Regge and Wheeler \cite{wheeler} demonstrated not the instability but the stability of the Schwarzschild black hole (for the full account see the classical text book of Chandrasekhar \cite{chand}). Following this a plethora of exact solutions, often mathematically discovered well before but not physically understood, were studied in their classical properties \cite{kerr}. Furthermore, numerous and quite restrictive mathematical theorems governing the existence as well as the geometric and topological properties of black holes were developed through the years (see for example \cite{maison}). Further interest was triggered by the work of Bekenstein and Hawking on quantum properties of black holes, in particular black hole thermodynamics. In recent years, research  for new black hole solutions has gradually developed in higher dimensions \cite{reallemparan} and/or for higher-dimensional extensions to general relativity \cite{lovelock}. This has been motivated by string theory and more recently by the brane world paradigm. Generically speaking, when we go to higher dimensions, novel permitted topologies and horizon geometries arise, such as the celebrated black ring \cite{blackring}, and the quite restrictive four-dimensional theorems are either no longer applicable or a lot less powerful (see for example \cite{ishibashi}).

One important characteristic of Einstein's field equations is that there exist non-trivial (ie non-maximally symmetric) solutions of the vacuum. The reason behind this fact is that, Einstein equations do not involve the Weyl tensor and as a result non-trivial geometric configurations (ie non-zero Riemannian curvature) can trigger non-trivial vacuum solutions. These vacuum solutions, at least in $d=4$, often turn out to be black hole solutions. Indeed assuming spherical symmetry immediately dictates the Schwarzschild black hole as the unique solution. As a result the gravitational field of spherical stellar objects is very well approximated by the far away field of this black hole solution. What happens when we include matter? The most simple inclusion of matter fields is that of an electric or magnetic charge. In this case Birkhoff's uniqueness theorem still holds since it has a clear analogue in Maxwell theory - a spherically symmetric electric charge does not radiate in the vacuum. The case of scalar fields, as matter or non-minimally coupled to gravity, are another obvious case to consider. Indeed, on the one  hand,  they are a natural and necessary ingredient of cosmology : a timelike scalar field  with potential is the correct way to describe a perfect fluid from a matter action principle. On the other hand, low energy effective actions resulting from string theories, include scalar fields as the dilaton or moduli associated with the size of extra dimensions (following the Kaluza-Klein paradigm \cite{KK}). Furthermore, an exponential or Liouville potential represents the higher-dimensional cosmological constant present in non-critical string theories \cite{nonsusy,moudas}, non-trivial curved adS backgrounds \cite{CLSZ}, or leading $g_s$ corrections to critical string theories in a flat background. In the case of scalars it is fair to say that they are not as "compatible" as Maxwell fields when coupled to general relativity. They are incompatible in the sense that they  can often spoil asymptotic properties, or again, render event horizons singular and as a result are not even permitted in the regularity hypothesis of no hair theorems (see for example \cite{bekenstein}). To illustrate a simple example consider Brans-Dicke theory, which turns out to have the same black hole solutions as ordinary GR-namely the scalar degree of freedom has to be  frozen in order to avoid a naked singularity. Moreover, since Birkhoff's theorem is no longer true (\cite{taub}, \cite{charm0}), scalars will on the other hand introduce star solutions differing from general relativity. They are heavily constrained  by local gravity tests, clearly favoring pure general relativity. To evade such problems one has to consider scalar potentials or couplings with other matter fields \cite{bekenstein}.

In this paper we will study in some detail the simplest non-trivial theories including a scalar and Maxwell field with a Liouville potential. Although this theory does not capture all the details of string effective actions (where additional fields are present, like the axion, see \cite{matos}, potentials are more complicated, etc.,) it contains the essential ingredients that will result in some of the phenomena we described above. Given the complicated theory at hand we will restrict ourselves to cylindrical symmetry or constant curvature horizons (in other words, $d-2$ homogeneous subspaces of constant curvature). Past work on this subject is rather scarce, when one includes the Liouville potential, since the difficulty in finding exact solutions is greatly enhanced. Indeed,  in the absence of the Liouville potential the problem is integrable as was demonstrated first by Gibbons and Maeda \cite{gibbonsmaeda} (see also \cite{horowitz} for a particular case and \cite{leygnac}). A  way to  give the dilaton a mass is with a Liouville potential, which can be derived from higher-dimensional string theory with a central deficit charge, for example (see also \cite{harveyhorohorne}).  A Liouville potential, or any potential for that fact, renders the problem of finding horizon covered singular solutions far more difficult technically. As we will see, even a pure cosmological constant is enough to spoil the integrability of the field equations. Furthermore, for a Liouville potential, asymptotics are non-trivial (in other words non-homogeneous spacetimes) since maximally symmetric solutions do not minimise the action \cite{wiltshire} or equivalently are simply not solutions to the field equations. This was also noted in the context of non-SUSY string theories \cite{nonsusy}. More precisely, in \cite{wiltshire},  using the theory of dynamical systems, Wiltshire and collaborators demonstrated that no regular asymptotic black holes exist in dilaton gravity, except in the case of a pure cosmological constant and an asymptotically frozen scalar field. Moreover, it was also shown that even in that case, no asymptotically de Sitter solutions exist ($\Lambda$>0).  These facts follow closely the no-hair theorems of general relativity. Based on these, Mann \emph{et al.} (\cite{CHM}) found and studied static spherically symmetric solutions for specific values of the coupling constants. They argued that these solutions are of physical interest as they have finite quasilocal charges and all terms in the action are finite outside the regular outer horizon. Unusual asymptotics were found in accord to Wiltshire's work as  a generic effect due to  the scalar potential not having a local minimum. Indeed in  the case of the Liouville potential we have a runaway minimum at infinite scalar field and this is enough  to guarantee good behaviour at infinity but still not maximal symmetry. The cylindrically symmetric problem in the absence of the EM field was studied in \cite{charm0} (see the work of Taub for the equivalent $\Lambda=0$ case). Closely after the work of \cite{CHM}, these black hole solutions were generalised to horizons of toroidal and hyperbolic geometry by Cai \emph{et al.} (\cite{Cai}), again for specific couplings. Several other works treated cases close to these \cite{rest}.

Application of the solutions we will find here are plentiful and do not only include usual general relativity. The backgrounds we will find can be applied to  a variety of differing problems and applications of high energy physics and cosmology. They will correspond to 5-dimensional backgrounds of black holes with a cosmological constant \cite{real}, cosmic p-brane extensions \cite{ruth} (see \cite{EMDsugra} for an extension to non-zero potential) and higher-codimension braneworld backgrounds \cite{roberto}. They also describe backgrounds for non-critical string theories as well as non-SUSY strings in 10 dimensions \cite{nonsusy}. We will analyse in all generality the integrability of cylindrically symmetric spacetimes and in particular constant curvature subspaces. We will establish an extension of the EM duality known for $\Lambda=0$ and discussed in a slightly different context in \cite{CLSZ}. We will then integrate down the system of coupled differential equations to 1 or 2 master equations depending on the symmetry at hand. With these generic results we will stick to 4 dimensions and find for general or special couplings exact solutions for the setup. For given coupling relations we will be able to write down  the general solution for the setup. This is particularily useful in order to understand the global properties of this system. The solutions we will find will include known and novel black hole and soliton solutions of the same symmetry and field content and can be easily extended in higher dimensions.

In the next section we will discuss the generic setup in $d$ dimensions and how to partially integrate the system. We will also establish solution-generating symmetries. We will give explicit solutions in the case of 4-dimensional spacetime, in section 3 for cylindrical symmetry and in section 4 for maximal symmetry. Novel general solutions, for given symmetry, and for families of coupling constants will be obtained and all known solutions (to the best of our knowledge) will be recovered. In section 5 we will consider the uplift of several of the solutions discussed in 4 dimensions to 5 and we will discuss solutions with squashed horizons \cite{squash}. We will conclude in the last section.

\section{Set-up and integrability}

Let us consider the $d$-dimensional action,
\beq
	\label{action}
	S = \int \ud^dx\sqrt{-g}\l[R-\frac12(\partial \Phi)^2-\frac14 e^{\ga\Phi}\mathcal F^2-2\Lambda e^{-\da\Phi}\r],
\eeq
where $\ga$, $\da$ and $\Lambda$ are coupling constants, $\mathcal F$ is the two-form field strength of the Maxwell one-form $A$, and $\Phi$ is the scalar field. This action covers, for different values of the coupling constants, several interesting cases. For example, given specific dimension-dependent values of
\beq
\label{KK}
 \ga=\pm\sqrt{\frac{2(d-1)}{d-2}},\qquad \da=\pm\sqrt{\frac{2}{(d-1)(d-2)}},
\eeq
theory (\ref{action}) is the Kaluza Klein reduction of  a $(d+1)$-dimensional general relativity with a cosmological constant and rotation or twist (see for example \cite{CLSZ}). The case $\delta=0$ reduces the Liouville potential to a cosmological constant and $\gamma=0$ to a pure Maxwell theory with a scalar kinetic term. In a completely different physical setting and taking $\mathcal F=0$ in $d=10$ dimensions, the action (\ref{action}) describes  tachyon-free non-supersymmetric string theory (\cite{moudas}, \cite{nonsusy}). The Liouville coupling $\gamma$ plays the role of the leading string surface  ($g_s$) correction  in the Liouville term which appears due to the breaking of supersymmetry. For example  we have $\ga=3/2$ for the type I string and $\ga=5/2$ for the closed heterotic string.  As we mentioned in the introduction, the characteristic of these string theories is that they do not have maximally symmetric backgrounds and as a result, the solutions of maximal possible symmetry are $SO(9)$ symmetric backgrounds \cite{moudas}.

The equations of motion stemming from the action take the form :
\bsea
	0&=&\partial_\mu\l(\sqrt{-g}e^{\ga\Phi}\mathcal{F}^{\mu\nu}\r), \slabel{Maxwell_modified} \\
	\square{\Phi}&=&\frac{\ga}4e^{\ga\Phi}\mathcal{F}^2-2\da\Lambda e^{-\da\Phi}, \slabel{Dilaton_modified} \\
	G_{\mu\nu}=\mathcal R_{\mu\nu}-\half\mathcal Rg_{\mu\nu}&=&\mathcal{T}_{\mu\nu} = \half\partial_\mu\Phi\partial_\nu\Phi-\frac{g_{\mu\nu}}4\l(\partial\Phi\r)^2+\half e^{\ga\Phi}\mathcal{F}^{\;\rho}_\mu\mathcal{F}_{\nu\rho}-\frac{g_{\mu\nu}}8e^{\ga\Phi}\mathcal F^2-\Lambda e^{-\da\Phi}g_{\mu\nu}, \slabel{Einstein_eq}
\esea
where $\square$ is the $d-$dimensional d'Alembertian, $G_{\mu\nu}$ the Enstein tensor and $\mathcal T_{\mu\nu}$ the stress-energy tensor, which has both an electromagnetic and a scalar contribution. This allows us to write the Ricci scalar, that is :
\beq
	\mathcal R = \frac{2}{2-d}\mathcal T=\half\l(\partial\Phi\r)^2+\frac{4-d}{4(2-d)}e^{\ga\Phi}\mathcal F^2-\frac{2d}{2-d}\Lambda e^{-\da\Phi}.
\eeq
The electromagnetic contribution vanishes as expected for $d=4$ (the Maxwell stress-energy tensor is traceless in 4 dimensions). Then, the only matter singular points of spacetime will be those present in the scalar field, see \cite{GHT}. However, for higher dimension, there might be a richer variety of singular points, though, in all the solutions we show in the following, all singular points of the Maxwell field are always contained in the dilaton field.

In this work we will consider a $d$-dimensional metric of the form (see also \cite{CLSZ}),
\beq
\label{cylindrical_metric}
	\ud s^2 = e^{2\chi}\al^{-\frac{d-3}{d-2}}(\ud r^2+ \ud \theta^2)+\al^\frac{2}{d-2}\l(-e^{2 U_t}dt^2+e^{2 U_\varphi} \si(\theta) ^2 \ud \varphi^2 +\sum_{i=1}^{i=d-4}e^{2U_i}\ud x^2\r),
\eeq
where the Maxwell field will be restricted to be either electric, $\mathcal A=A(r,\theta)dt$ or magnetic $\mathcal A=A(r,\theta)d\phi$ (for dyonic solutions see \cite{bulg}). The function $\si(\theta)$ denotes $\sin(\theta)$, $\sinh(\theta)$ and unity for $\kappa=1,-1,0$ respectively. We can also choose the potentials $U_i$ so that  they sum to zero,
\beq
\label{sum}
\sum_{i=1}^{i=d-4}U_i+U_t+U_\varphi=0
\eeq
without any loss of generality.
When $\kappa=0$ and all metric components are locally only $r$-dependent we have cylindrical symmetry ($r$ is not the normal coordinate). For $d=4$, $\kappa=\pm1$ will correspond  to a spherically symmetric and hyperbolic $2$-dimensional spacelike sections respectively {\footnote{There is no particular reason in choosing 2-dimensional sections for a $d$-dimensional spacetime except that in the present analysis we will specialize later on to 4-dimensional spacetimes. This can be easily generalised \cite{ruth1}}}.
It is rather useful now to go to a different set of variables \cite{CLSZ} for which the field equations will take a simpler form,
\bea
	\psi_\star& = & \sqrt\frac{d-2}{d-3}\l[\frac{d-3}{d-2}(\Phi -\da\ln \al)+ \ga U_\star\r],	\slabel{chgt1}\\
	\psi_i& =& U_i+\frac{1}{d-3}U_\star,\qquad i=1,...,d-4\slabel{chgt12}\\	
	\Omega & = & \ga(\Phi-\da\ln\al)-2U_\star,	\slabel{chgt2}\\
	2\nu & = & 2\chi -\da \Phi +\frac{\da^2}2 \ln\al.	\slabel{chgt3}
\eea
where $\epsilon=-1$ corresponds to an electric potential and $\epsilon=1$ magnetic one. The $\star$ symbol denotes  $t$ for the electric case and $\varphi$ for the magnetic case respectively.
These technicalities put aside, the field equations for the electric case ($\eps=-1$) take the form,
\bsea
	\al'' -\kappa \al & = & -2\Lambda \al^{\frac{1}{d-2}-\frac{\da^2}{2}}e^{2\nu} \slabel{m1}\\
	0 & = & \overrightarrow{\nabla} \cdot \l(e^{\Omega} \al^{\ga\da+\frac{d-4}{d-2}} \si(\theta)^{-\eps} \overrightarrow{\nabla} A  \r), \slabel{m2}\\
	\l(\al \Omega'\r)' +\Big(\ga\da-\frac{2}{d-2}\Big)\alpha \kappa & = & \dfrac{\eps s}2 e^\Omega \al^{\ga\da+\frac{d-4}{d-2}} \si(\theta)^{-\eps}\l( \overrightarrow{\nabla} A\r)^2 ,	\slabel{m3}\\
	\l(\al \psi'_\star\r)' +\alpha \kappa \sqrt{\frac{d-3}{d-2}}\Big(\da+\frac{\ga}{d-3}\Big) &=& 0,	\slabel{m4}\\
	\l(\al \psi_i'\r)' & = & 0,	\qquad i=1,..,d-4 \slabel{m5}\\
	2\nu'\dfrac{\al'}\al-\dfrac{\al''}\al -\kappa & = & \dfrac1s\l((\psi_\star')^2+\dfrac12(\Omega')^2\r)+\dfrac\eps2 e^\Omega \al^{\ga\da -\frac{2}{d-2}} (A'^2-\dot{A}^2)+\sum_{i=1}^{d-4} 		\psi_{i}^{'2} ,	\slabel{m6} \\
	2\al\nu' \kappa -\Big(\frac{d-1}{d-2}-\frac{\da^2}{2}\Big)\al'\kappa &=&\frac{2\kappa \alpha}{s}\l[-\Big(\frac{\ga}{d-3}+\da\Big)\sqrt{\frac{d-3}{d-2}}\psi'-\dfrac12 \Big(\ga\da-\frac{2}{d-2}\Big)\Omega'\r] \slabel{m7},	
\esea
All fields depend on $r$ (according to cylindrical symmetry) except the electric potential for which we allow a $(r,\theta)$ dependence which will be useful for the extension of the electro-magnetic duality in 4 dimensions later on. For the same reason we keep $\epsilon$. Note equation (\ref{m7}) which is an additional equation present for $\kappa \neq 0 $ which constrains the metric elements (\ref{cylindrical_metric}) in such a way as to obtain maximally symmetric 2-dimensional sections. We have also set
\beq
s=\gamma^2+2\frac{d-3}{d-2}
\eeq
For the magnetic case ($\eps=1$) on the other hand we have,
\bsea
	\al'' -\kappa \al& = & -2\Lambda \al^{\frac{1}{d-2}-\frac{\da^2}{2}}e^{2\nu},	\slabel{m11}\\
	0 & = & \overrightarrow{\nabla} \cdot \l(e^\Omega\si{\theta}^{-\eps} \al^{{\ga\da}+\frac{d-4}{d-2}}\overrightarrow{\nabla} A \r), \slabel{m12}\\
	\l(\al \Omega'\r)' +\Big(\ga\da+\frac{2(d-3)}{d-2}\Big)\alpha \kappa & = & \dfrac{\eps s}{2}\si(\theta)^{-\eps} e^\Omega \al^{\ga\da+\frac{d-4}{d-2}} \l( \overrightarrow{\nabla} A\r)^2,	\slabel{m13}\\
	\l(\al \psi_\star'\r)' +\alpha \kappa \sqrt{\frac{d-3}{d-2}}(\da-\ga)& = & 0,	\slabel{m14}\\
	\l(\al \psi_i'\r)' & = & 0,	\qquad i=1,..,d-4\slabel{m15}\\
	2\nu'\dfrac{\al'}\al-\dfrac{\al''}\al -\kappa &=& \dfrac1s\l((\psi')^2+\dfrac12(\Omega')^2\r)+\frac{\eps}{2\si^2(\theta)}e^\Omega \al^{\ga\da -\frac{2}{d-2}} (A'^2-\dot{A}^2) \nonumber\\
	&&+\sum_{i=1}^{d-4}\psi_i^{'2} ,	\slabel{m16} \\
	2\al\nu' \kappa- \Big(\frac{d-1}{d-2}-\frac{\da^2}{2}\Big)\al'\kappa&=&\frac{2\alpha \kappa}{s}\l[(\ga-\da)\sqrt{\frac{d-3}{d-2}}\psi'-\dfrac12 \Big(\ga\da+2\frac{d-3}{d-2}\Big)\Omega'\r],\slabel{m17}	
\esea

The field equations written in this form are quite straightforward to reduce to one or two coupled second order ODE's with respect to one or two  variables respectively. In reducing the system of equations, we adapt our system of coordinates accordingly. It turns out that the judicious system of coordinates differs for $\kappa=0$ (cylindrical symmetry) and for $\kappa\neq 0$. Let us reduce the system in turn now for each case, starting with $\kappa=0$. Note that (\ref{m7}) and (\ref{m17}) drop out in this case.

\subsection{Case of Cylindrical symmetry}

Directly integrating \eqref{m2}, \eqref{m4}, \eqref{m5} and using \eqref{m3}, we get
\bsea
Q&=&e^\Omega\al^{\ga\da+\frac{d-4}{d-2}}A'\slabel{Q}\\
\al \Omega' &=& \dfrac{s\eps}2 QA +a \slabel{A}\\ 
c_t & = & \al \psi'_t \slabel{c_r},\qquad c_i = \al \psi'_i \slabel{c_i}
\esea
where $Q$ is the electric charge and $c_t, c_i$ are the constant scalar charges associated to the $\psi$ fields. The constant $a$ can be gauged away but we choose to keep it and fix it later to simplify integrated quantities.
Using now (\ref{m1}), (\ref{m6}) and (\ref{A}), we solve for  $\nu'$
\beq
2\nu'\al = \l( \frac{d-1}{d-2} -\frac{\da^2}{2}\r)\al' + \l(\frac{2}{d-2}-\ga\da\r)\l(\frac{\eps}{2}QA+\frac as\r)-h,
\label{h_r}
\eeq
Here $h$ is a constant and can be related with the energy/quasilocal mass of the solution.
At the end of the day the whole system boils down to 2 coupled ODE's,
\bsea
	 \dfrac{s\eps}2 QA + a + \l( \ga\da+\frac{d-4}{d-2} \r)\al' +\al \dfrac{A''}{A'}&=&0, \slabel{alphaA1}\\
	\al'  \l[ \l( \frac{d-1}{d-2}-\frac{\da^2}{2} \r)\al'+\l(\frac{2}{d-2}-\ga\da \r)\l(\frac as +\frac{\eps A Q}{2} - h\r)\r] - \al'' \al &=& \dfrac1s \l[c_\star^2+\dfrac12\l(\dfrac{s\eps}2 QA+a\r)^2\r] \nonumber\\
	&+&\frac{\eps Q}{2} \al A' +\sum c_i^2, \slabel{alphaA2}
\esea
which once solved give a solution for theory (\ref{action}) with cylindrical symmetry (\ref{cylindrical_metric}). To go further we fix the coordinate system by setting $\al' = p$.
The integration of equation \eqref{alphaA1} then gives :
\beq
k = \frac{s\eps Q}4A^2 +aA +\al A' - \l(\frac{2}{d-2}-\ga\da\r)\l(pA-\int A\,\ud p\r),
\label{kr}
\eeq
On the other hand, \eqref{alphaA2} then becomes
\bea
\label{X}
X(p)&\equiv& \l( \frac{d-1}{d-2} -\frac{\da^2}{2}\r)p^2 +\l[\l(\frac{2}{d-2}-\ga\da \r)\frac as - h\r]p -\frac{\eps Q}2 k-\frac{c^2}s -\frac{a^2}{2s} \nonumber \\
&=& - \frac{Q\eps}2\l(  \frac{2}{d-2}-\ga\da \r)\int A\,\ud p +p\frac{\ud p}{\ud \ln\al}.
\eea
Let us note now that taking $\frac{2}{d-2}-\ga\da=0$ \eqref{X}  completely decouples from $A$ and gives immediately $\alpha$. The Kaluza-Klein case falls in this category, see \eqref{KK}. Once $\alpha$ is known, $A$ is obtained from \eqref{kr}. We will examine shortly and in detail the solutions emanating for the 4-dimensional case. Lastly, by defining
\beq
	B(p)=\int A(p)\ud p
\eeq
and combining \eqref{X} and \eqref{kr} we get
\beq
	k-\frac{s\eps Q}4\dot B^2 -a\dot B + \l( \frac{2}{d-2}-\ga\da \r)(p\dot B -B) = \ddot B\l[X(p)+  \frac{Q\eps}2\l(  \frac{2}{d-2}-\ga\da \r)B\r].
	\label{radial_eq}
\eeq
This second-order, non-linear and autonomous ODE with respect to $B$ is one of the  main results in this section. Once we have determined $B$ analytically or numerically we can then find a solution of the entire system which corresponds to an exact solution of  the action (\ref{action}) for a metric of cylindrical symmetry (\ref{cylindrical_metric}). We will find several solutions in the next sections for the case of 4 dimensions. Indeed, once $B$ is known from \eqref{X}, it is easy to see that
\beq
	\ln \alpha(p)= \int \frac{p\,\ud p}{X(p)- \frac{Q\eps}2(\ga\da-\frac2{d-2})B}.
	\label{logalpha}
\eeq
Using \eqref{h_r}, we solve for $\nu$
\beq
	2\nu=\l( \frac{d-1}{d-2} -\frac{\da^2}{2}\r)\ln\alpha +\int \frac{\ud p}{X(p)- \frac{Q\eps}2(\ga\da-\frac2{d-2})B} \l[-h+\l(  \frac{2}{d-2}-\ga\da  \r)\l(\frac{\eps}{2}QA+\frac as\r)\r].
	\label{nu_eq}
\eeq
Note that alternatively, equation \eqref{m1} enables us to write :
\beq
	e^{2\nu}=-\frac{1}{2\Lambda}\l[X(p)+ \frac{Q\eps}2\l(   \frac{2}{d-2}-\ga\da \r)B\r]\alpha^{-\frac{d-1}{d-2}+\da^2/2}.
	\label{solnu}
\eeq
These two equations fix $\Lambda$ with respect to the integration constants.
We make use of \eqref{c_r} to write
\beq
	\psi = c\int \frac{\ud p}{X(p)+ \frac{Q\eps}2\l( \frac{2}{d-2}-\ga\da  \r)B}.
	\label{solpsi}
\eeq
and similarly for $\psi_i$. Finally to get $\Omega$ we use \eqref{Q}
\beq
	e^{\Omega}= \frac{Q\al^{\frac{2}{d-2}-\ga\da}}{\dot A(p) \l[X(p)+ \frac{Q\eps}2\l( \frac{2}{d-2}-\ga\da  \r)B\r]}.
	\label{solOmega}
\eeq
Note that, in terms of $p$, the line element becomes :
\beq
	\label{cylindrical_metric_bis}
	\ud s^2 =\frac{\al^{\frac{d-1}{d-2}}e^{2\chi}\ud p^2}{\l[X(p)+\frac{Q\eps}2\l(  \frac{2}{d-2}-\ga\da \r)B\r]^2} + e^{2\chi}\al^{-\frac{d-3}{d-2}}\ud z^2+\al(-e^{2\U}\ud t^2+e^{-2\U}\,\ud\varphi^2).
\eeq
It is important to note here that (\ref{solnu}) and (\ref{nu_eq}) provide a relation between the action parameter $\Lambda$ and the constants of integration. This is similar to the pure dilatonic case \cite{charm0}. Here (\ref{solOmega}) and (\ref{Q}) provide an additional relation between constants of integration.

\subsection{Maximally symmetric case}

We now turn our attention to the case of $\kappa\neq 0$. Let's stick to the electric case here and note that the judicious choice of coordinates dictated for example from \eqref{m4} is now
\beq
	\al=\frac{\ud x}{\ud r}.
\eeq
The two coordinate systems are related by $2p=\dot{\al^2}$. Hence, if $\al^2$ is a second degree polynomial only then are $p$ and $x$ identical coordinates.
We also note, for the electric case, that in order for the $A$ field not to be trivial - imposed by separability requirements -, it has to be a function of $r$. On the other hand for the magnetic case $A$ has to be a function of $\theta$.  This can easily be seen by inspecting the equations of motion in both these cases. Other than that the magnetic resolution is very similar to the electric one.
Let's denote by a dot the derivation with respect to $x$. Integrating \eqref{m2}, \eqref{m3} and then \eqref{m4}, \eqref{m5} we obtain
\bsea
	Q&=&e^\Omega\al^{\ga\da+2\frac{d-3}{d-2}}\dot A, \slabel{Q_max} \\
	\al^2\dot\Omega &=& \eps \frac s2QA +a +\l(\frac{2}{d-2}-\ga\da\r)\ka x, \slabel{A_r_max}\\
	\al^2\dot\psi &=&c- \ka x \sqrt{\frac{d-3}{d-2}}\l(\da+\frac{\ga}{d-3}\r), \slabel{c_x_max}\\
	\al^2\dot\psi_i&=& c_i  \slabel{ci_x_max}.
\esea
Combining \eqref{m6} and \eqref{m7} with \eqref{A_r_max}, \eqref{c_x_max} we obtain
\beq
	2\al^2\dot\nu = \l[\frac{2}{(d-2)(d-3)}+\da^2\r]\ka x+ \l(\frac{d-1}{d-2}-\frac{\delta^2}{2}\r)\al\dot\al+\l(\ga\da-\frac{2}{d-2}\r)\l(\frac{a}{s}-\frac{QA}{2}\r)-h.
	\label{h_x_max}
\eeq
where we now have
\beq
\label{hconstraint}
	h=\frac{2c}{s}\sqrt{\frac{d-3}{d-2}}\l(\delta+\frac{\gamma}{d-3}\r)
\eeq
so that  \eqref{m6}, \eqref{m7} are compatible - maximal symmetry imposes one more  relation between the integration constants $h$ and $c$. In fact for $\kappa=0$ this means that the $(z,\phi)$ plane is homogeneous in the cylindrical case (\ref{cylindrical_metric_bis}). We have now solved the system with respect to the variables $\al$ and $A$. Indeed using \eqref{A_r_max} and \eqref{m2} we obtain
\beq
\label{int1}
	\al^2 \frac{\ddot{A}}{\dot{A}}+2\dot{\al}\al-\l(\ga\da-\frac{2}{d-2}\r)(\ka x-\dot{\al}\al)+a-\frac{sQ A}{2}=0
\eeq
and then using \eqref{h_x_max} with \eqref{m6} we get
\bea
\label{int2}
	&&\dot{\al} \al\left[\l(\delta^2 +\frac{2}{(d-2)(d-3)}\r)\kappa x \right.+ \left.\frac{\dot{\al}\al}{2}\l(\frac{2}{d-2}-\da^2\r)\right] + \l(\ga\da-\frac{2}{d-2}\r)\l(\frac{a}{s}-\frac{QA}{2}\r)(-\dot{\al}\al+\kappa x)- \nonumber \\ &&- h(\dot{\al}\al-\kappa x)
=\ddot{\al}\al^3+\kappa \al^2+ \frac{c^2}{s}+\kappa^2x^2\l(\frac{\da^2}{2}+\frac{1}{(d-2)(d-3)}\r) + \frac{1}{2s}\l(a-\frac{sQA}{2}\r)^2-\frac{Q\alpha^2\dot{A}}{2}
\eea
By solving for $A$ and $\al$ we find a solution to the full system \eqref{m1}-\eqref{m7} by direct integration of \eqref{A_r_max}-\eqref{h_x_max}. In particular note that for $\ga\da=\frac{2}{d-2}$ (\ref{int1}) integrates out giving
\beq
Q e^{-\Omega}=\al^2 \dot{A}= k-a A-\frac{sQ A^2}{4}.
\eeq
where we have also used (\ref{Q_max}) to obtain $\Omega$. This reduces the full system to the resolution of equation (\ref{int2}) with respect to $A$.

This completes our analysis of the theory  in arbitrary dimension $d$. From now on we will concentrate on the case of $d=4$, describing a symmetry of the equations of motion and then giving explicit solutions as well as their uplifted counterparts.

\subsection{Electro-magnetic duality in $d=4$}

Let us consider now the symmetries of the magnetic and electric field equations \eqref{m1}-\eqref{m7} and \eqref{m11}-\eqref{m17} (we follow \cite{CLSZ}). We can define a dual potential $\omega$ to $A$ through
\beq
\label{dualpotential}
\l(-\partial_\theta \omega, \partial_r \omega\r)=e^\Omega \al^{\ga\da} si(\theta)^{-\epsilon}\l(\partial_r A, \partial_\theta A\r)
\eeq
To be definite we take $\epsilon=-1$ and apply \eqref{dualpotential}. Upon doing this the field equations \eqref{m2}, \eqref{m3}, \eqref{m6} and \eqref{m7} take the form,
\bsea
	0 & = & \overrightarrow{\nabla} \cdot \l(e^{-\Omega} \al^{-\ga\da} (\si(\theta))^{-1} \overrightarrow{\nabla} \omega  \r), \slabel{m2d}\\
	\l(\al \Omega'\r)' +\l(\ga\da-1\r)\alpha \kappa & = & \dfrac{\eps s}2 e^{-\Omega} \al^{-\ga\da} (si(\theta))^{-1}\l( \overrightarrow{\nabla} \omega\r)^2 ,	\slabel{m3d}\\
	2\nu'\dfrac{\al'}\al-\dfrac{\al''}\al -\kappa & = & \dfrac1s\l((\psi_\star')^2+\dfrac12(\Omega')^2\r)+\dfrac\eps2 e^{-\Omega} \al^{-\ga\da -1} (\omega'^2-\dot{\omega}^2) \nonumber\\
	&&+\sum_{i=1}^{d-4} \psi_{i}^{'2} ,	 \slabel{m6d} \\
	2\al\nu' \kappa -\l(\frac{3-\da^2}{2}\r)\al'\kappa &=&\frac{2\kappa \alpha}{s}\l[-\l(\ga+\da\r)\frac{\sqrt{2}}2\psi'-\dfrac12 \l(\ga\da-1\r)\Omega'\r] \slabel{m7d},	
\esea
Now consider the following map,
\beq
\label{dualitymap}
\bar{\Omega}=-\Omega,\qquad \bar{A}=\omega, \qquad \bar{\epsilon}=-\epsilon,\qquad \bar{\gamma}=-\ga,\qquad \bar{\da}=\da,
\eeq
then (\ref{m2d}), (\ref{m3d}), (\ref{m6d}) and (\ref{m7d}) are exactly (\ref{m12}), (\ref{m13}), (\ref{m16}) and (\ref{m17})
 for the barred variables $\bar{A}, \bar{\Omega}$ and constants $\bar{\ga},\bar{\da},\bar{\epsilon}$. The remaining equations (\ref{m11}), (\ref{m14}), (\ref{m15}), (\ref{m17}) do not yield any additional constraint and hence the map (\ref{dualitymap}) generates a novel solution. In other words, the duality is valid for any $\ga$ and $\da$. The application (\ref{dualitymap}) provides a simple way to obtain a magnetic/electric solution from another given electric/magnetic solution. Although (\ref{dualitymap}) is clearly an extension of the EM duality for $\Lambda=0$ it is of quite a different nature since (\ref{dualitymap}) changes the coupling $\ga$, hence maps solutions belonging to \emph{different} theories. We will use this symmetry in order to construct solutions in $d=4$ dimensions for $\kappa\neq 0$. This will also be particularly useful to construct solutions for the uplifted metrics.

\section{Cylindrical solutions in 4-dimensional spacetime}

In all generality we need only to solve (\ref{radial_eq}) which is not integrable in general. There are, however, several special cases depending on the coupling constants of our theory $\ga$ and $\da$. In fact it is easy to see that (\ref{kr}) and (\ref{X}) are decoupled when $\ga \da=1$ and in this particular case we can obtain the general solution. We deal with this case first.

\subsection{The general solution for $\ga \da=1$}

	\label{subsubsection:ka0_general}

Combining \eqref{kr} and \eqref{X} we obtain
\beq
\label{cond}
	\frac{\ud A}{-\frac{\eps sQ}4A^2 -aA+1} = \frac{\ud p}{\frac{3-\da^2}2 p^2 - hp -\frac{\eps Q}2 -\frac{c^2}s -\frac{a^2}{2s} },
\eeq
where we have rescaled $k$ and we remind the reader that the constant $a$ is arbitrary, reflecting a choice of coordinates which we now fix.
We demand the discriminants of both polynomials to be equal to each other and positive. Hence we set
\beq
	\Delta_X=\Delta_A = a^2+\eps sQ = \la^2 > 0,
\eeq
where $\la$ is now arbitrary replacing $a$, hence
\beq
	X(p) = \frac{3-\da^2}2 p^2 - hp -\frac{c^2}s -\frac{\la^2}{2s}.
	\label{X:1}
\eeq
The discriminant $\Delta_X$ on the other hand is always positive for $\da^2<3$. When $\da^2>3$ we need to suppose additionally that $h^2>\frac{2c^2}{s} (\da^2-3)$\footnote{The case $h^2\leq \frac{2c^2}{s} (\da^2-3)$ can be dealt in a different coordinate system but does not present interesting black hole solutions.}. The case of $\delta^2=3$ is also special and will be treated later. First, let us  take $\da^2<3$
and then proceed to a re-scaling of coordinates
\beq
	\bar h =\frac h{|\la|}, \quad \bar c= \frac c{|\la|}, \quad \bar p = \frac{3-\da^2}{|\la|}p -\frac h{|\la|}+1.
\eeq
Dropping anew all the bars and comparing with (\ref{X}) we have
\beq
	h^2+(3-\da^2)\frac{2c^2}s = \frac{(\da^2-1)^2}{\da^2+1},
	\label{rel_h-c-lambda:1}
\eeq
which imposes certain conditions on $h$ and $c$. In particular, we see that the case $\ga=\da=1$ has to be treated separately and will be dealt with after this section.
We now integrate (\ref{cond})
\beq
	\dot A(p) = 2e^{-\frac{\Phi_0}{2\da}}\sqrt{\frac{\da^2(1-\eta^2)}{1+\da^2}}\frac1{(\eta p +1 -\eta)^2}, \quad X(p)=\frac{\la^2}{2(3-\da^2)}p(p-2),
	\label{A:1}
\eeq
where the dot denotes derivation with respect to $p$, the EM charge has been replaced by its expression in terms of $\Phi_0$, an integration constant linked to the dilaton field and the integration constant $\eta$ is such that $|\eta|<1$. The zeros of $X$, $p=0$, $p=2$ and the singularity in $A$, $p_\eta=1-1/\eta$ will be possible singularities or horizon positions for the metric. We will call them singular points for reference.

Using the integrals obtained in the previous section we can now write down the general solution for the case of cylindrical symmetry :

\bea
	\ud s^2 &=& -(p-2)^{C(-h,-c,\eps)}p^{C(h,c,\eps)}(\eta p+1-\eta)^{\frac{2\eps\da^2}{1+\da^2}} \ud t^2 +\nonumber\\
			&&\frac{e^{\da\Phi_0}}{-\Lambda(3-\da^2)}p^{F(h,c)}(p-2)^{F(-h,-c)}(\eta p+1-\eta)^{\frac{2\da^2}{1+\da^2}}\ud p^2 +\nonumber \\
			&& +p^{B(h,c)} (p-2)^{B(-h,-c)}(\eta p+1-\eta)^{\frac{2\da^2}{1+\da^2}} \ud z^2 \nonumber \\
			&& + p^{C(h,c,-\eps)} (p-2)^{C(-h,-c,-\eps)} (\eta p+1-\eta)^{-\frac{2\eps\da^2}{1+\da^2}} \ud\varphi^2,
\label{metric:1}
\eea
and dilaton field
\beq
	e^{\Phi} = e^{\Phi_0}(\eta p+1-\eta)^{\frac{2\da}{\da^2+1}}p^{D(h,c)}(p-2)^{D(-h,-c)},
      \label{Phi:1}
\eeq
where we have rescaled the $t$, $z$ and $\varphi$ coordinates to absorb constant overall factors and
where the exponents are given by,
\bsea
\label{exp}
	F(h,c) &=& -1+\frac{\da^2}{3-\da^2}(1-h)-\da^2\frac{1+\da\sqrt2c}{1+\da^2}, \\
	B(h,c) &=&1+\frac{\da^2-2}{3-\da^2}(1-h)-\da^2\frac{1+\da\sqrt2c}{\da^2+1},\\
	C(h,c,\eps) &=& \frac{1-h}{3-\da^2}-\eps\da\frac{\da-\sqrt2c}{\da^2+1},\\
	D(h,c)&=&\frac{\da}{3-\da^2}(1-h)-\da\frac{1+\da\sqrt2c}{1+\da^2}.
\esea
Note the symmetry upon exchanging the sign of $h$ and $c$ and interchanging $p$ and $p-2$ :
\beq
	\label{symmetry}
	\ud s^2(h,c,p>2)=\ud s^2(-h,-c,p<0).
\eeq
Therefore we only need to study the $p>2$ interval, for which the coordinate $p$ is space-like iff
\beq
	-\Lambda(3-\da^2)>0,
\eeq
that is, staticity links the sign of $\Lambda$ and the value of $\da$. Let us take without loss of generality $p_\eta<0$, in other words $-1<\eta<0$. Summing it up, $p>2$ is spacelike if $\da^2<3$ ($\da^2>3$) and $\Lambda<0$ ($\Lambda>0$), resulting in an adS-like (dS-like) spacetime. For $\delta^2<3$ and $\Lambda<0$, the coordinate $p$ is timelike in between $0<p<2$ (the normal time coordinate is in fact infinite in this case), where of course $t$ in (\ref{metric:1}) is Wick rotated accordingly. Likewise, if $\delta^2<3$ and $\Lambda>0$, the coordinate $p$ is spacelike in between $0<p<2$, etc. All in all, after fixing $\da$ and $\Lambda$ \eqref{metric:1} represents 3 different solutions, 1 cosmological and 2 static (given (\ref{symmetry})).

On the other hand, for large $p$, note that $A$ flows to a constant, the boundary of the solution is conformally flat and the spacetime picks up $SO(1,2)$ symmetry,
\beq
ds^2\sim q^{2}(-dt^2+d\phi^2+dz^2)+\frac{dq^2}{q^{2(1-\da^2)}} \frac{e^{\da\Phi_0}}{(-\Lambda)(3-\da^2)}\eta^{\frac{2\da^2}{1+\da^2}}, \quad q=p^{\frac1{\da^2-3}}
\label{metric_asymptotic:1}
\eeq
This is  a Poincar\'e patch of adS space iff $\delta=0$ and $\Lambda<0$ (de Sitter for $\Lambda>0$). This is actually not surprising since in this class of solutions, $\ga\da$ equals $1$, which necessarily means that $\ga\rightarrow \infty$ accordingly. In fact, it is relatively easy to see, that in all generality, the limit $\da=0$ corresponds precisely to freezing the EM potential, which is evident by \eqref{A:1}. All spacelike uplifted solutions ($\da^2=\frac13$) in $d=5$ belong to this class and will only be valid for a negative cosmological constant.

To determine which points of space-time are singular, we need to calculate the Ricci scalar which can be done in two (equivalent) ways : either from the metric (41) or from the expression of the trace of the stress-energy tensor,
\beq
	\mathcal R = - \mathcal T = \half\l(\partial\Phi\r)^2 + 4\Lambda e^{-\da\Phi} =- \mathcal{T}_1-\mathcal{T}_2.
	\label{Ricciscalar_T}
\eeq
The expressions for $\mathcal{T}_1$ and $\mathcal{T}_2$ are :
\bsea
	\mathcal{T}_1 &=& \frac{2\Lambda\da^2e^{-\da\Phi_0}}{\l(\da^2-3\r)\l(1+\da^2\r)}\l(\eta p+1-\eta\r)^{-2-\frac{2\da^2}{\da^2+1}}p^{-1-\da D(h,c)}(p-2)^{-1-\da D(-h,-c)} \nn\\
&&\Bigg\{-(1+\da^2)p^2+\l[-(1+\da^2)\eta h+\eta(\da^2-3)\da\sqrt2c+2(1+\da^2)\eta+2(1-\da^2)\r]p \nn\\
&&+(\eta-1)\l[(1+\da^2)h-(\da^2-3)\da\sqrt{2}c-2(1-\da^2)\r]\Bigg\}^2,\\
	\mathcal{T}_2&=&4\Lambda e^{-\da\Phi_0}\l(\eta p+1-\eta\r)^{-\frac{2\da^2}{\da^2+1}}p^{-\da D(h,c)}(p-2)^{-\da D(-h,-c)},
\esea
so that formally
\beq
	\mathcal R = P_4\l(p,\eta,h,c,\da\r)\l(\eta p+1-\eta\r)^{-2-\frac{2\da^2}{\da^2+1}}p^{-1-\da D(h,c)}(p-2)^{-1-\da D(-h,-c)}.
	\label{Ricciscalar:1}
\eeq
We can immediately check that we recover $\mathcal R=4\Lambda$ in the $\da=0$ limit, in agreement with the remarks above.

Let us first look at the asymptotics $p\to\infty$ :
\beq
	\mathcal R\quad \sim_\infty \quad p^{\frac{2}{\da^2-3}}.
\eeq
Thus, the Ricci curvature will be regular as $p\to\infty$ iff $\da^2<3$, while it will diverge if $\da^2>3$.  Computation of the Weyl square using the asymptotic form \eqref{metric_asymptotic:1} of the metric yields the same behaviour for the Weyl square or the Kretschmann scalar. Together with the precedent remarks about staticity for $p>2$, this suggests that the coordinate system we are using is not adapted to a spacelike distance for the $\da^2>3$, $\Lambda>0$ case and we should change for $q=\frac1p$, $q>0$ whereupon $q=0$ is singular.

Given (\ref{Ricciscalar:1}) we see that there is always a curvature singularity at $p_\eta$ but it is more subtle  to read what happens at $p=0$ or $p=2$, which can be curvature singularities or horizon positions. Indeed, one can look at the sign of the exponent $-1-D(-h,-c)$. This is a function of two variables, $h$ and $\da$, since $c$ is constrained by \eqref{rel_h-c-lambda:1}. Plotting $-1-D(-h,-c)$ in terms of $h$ and $c$ shows that it is always negative. Computing the partial derivatives of this function with respect to $h$ and $\da$, we find a single extremum at $h=\frac{1-\da^2}2$, for which $D(-h,-c)=0$. Thus, except in the case where $h=\frac{1-\da^2}2$, $p=0$ and $p=2$ will be curvature singularities.

Now, using the freedom we have in $h$ and $\da$, we will try to regularize the solutions for $p=2$. The following statements are equivalent :
\begin{enumerate}
\item  The dilaton field is regular at $p=2$ $\Longrightarrow$ $D(-h,-c)=0$.
\item The Ricci scalar is regular at $p=2$ $\Longrightarrow$ $-1-D(-h,-c)=-1$ and $P_4\l(p,\eta,h,c,\da\r)=(p-2)P_3\l(p,\eta,\da\r)$.
\item $C(-h,-c,-1)=1$.
\item $F(-h,-c)=-1$.
\item $B(\pm h,\pm c)=C(\pm h, \pm c, 1)$.
\item $h=\da\sqrt2c\Longrightarrow h=\frac{1-\da^2}2$.
\end{enumerate}
Let us examine these regular solutions, discriminating between the $\eps=-1$ electric case and the $\eps=1$ magnetic case.

\subsubsection{Black Hole and regular solutions}
	\label{subsubsection:ka_0_generalBH}

Fixing $\eps=-1$ and $h$ as stated above, we obtain the following solution :
\bsea
	\dot A(p) &=& 2e^{-\frac{\Phi_0}{2\da}}\sqrt{\frac{\da^2(1-\eta^2)}{1+\da^2}}\frac1{(\eta p +1 -\eta)^2}, \slabel{A:2}\\
	e^{\Phi} &=& e^{\Phi_0}(\eta p+1-\eta)^{\frac{2\da}{\da^2+1}}p^{\frac{4\da(\da^2-1)}{(\da^2+1)(3-\da^2)}}, \slabel{Phi:2}\\
	\ud s^2 &=& -(p-2)\frac{p^{\frac{-\da^4+6\da^2-1}{(1+\da^2)(3-\da^2)}}}{(\eta p+1-\eta)^{\frac{2\da^2}{1+\da^2}}} \ud t^2 -\frac{e^{\da\Phi_0}}{\Lambda(3-\da^2)}\frac{p^{\frac{5\da^4-6\da^2-3}{(1+\da^2)(3-\da^2)}}}{p-2}(\eta p+1-\eta)^{\frac{2\da^2}{1+\da^2}}\ud p^2 \nonumber \\
			&& + \,p^{\frac{2(\da^2-1)^2}{(\da^2+1)(3-\da^2)}} (\eta p+1-\eta)^{\frac{2\da^2}{1+\da^2}}\big( \ud z^2 + \ud \varphi^2 \big),\slabel{metric:2}
	\label{solution:2}
\esea
$\eta$ is constrained to be $|\eta|<1$, which implies that $p_\eta<0$ ($\eta>0$) or $p_\eta>2$ ($\eta<0$). We will of course arrange for the first eventuality.

The Ricci scalar becomes :
\bea
	\mathcal R &=& - \mathcal{T}_1-\mathcal{T}_2=P_3(p,\eta,\da)\l(\eta p-\eta p_\eta\r)^{-2\frac{2\da^2+1}{\da^2+1}}p^{\frac{3\da^4-2\da^2+3}{(\da^2+1)(\da^2-3)}}  \\
	\mathcal{T}_1 &=& \frac{2\da^2\Lambda}{(1+\da^2)(3-\da^2)}e^{-\da\Phi_0}\l(p-2\r)\l[(\da^2+1)\eta p +2\l(\da^2-1\r)\l(1-\eta\r) \r]^2\l(\eta p-\eta p_\eta\r)^{-2\frac{2\da^2+1}{\da^2+1}}p^{\frac{3\da^4-2\da^2+3}{(\da^2+1)(\da^2-3)}}  \nn \\
	\mathcal{T}_2 &=&-4\Lambda e^{-\da\Phi_0}p\l(\eta p -\eta p_\eta\r)^2\l(\eta p -\eta p_\eta\r)^{-2\frac{2\da^2+1}{\da^2+1}}p^{\frac{3\da^4-2\da^2+3}{(\da^2+1)(\da^2-3)}} \nn
	\label{Ricciscalar:2}
\eea
which as expected is regular at $p=2$ and also when $p\to\infty$ iff $\da^2<3$. The curvature singularities displayed in the Ricci scalar can only be those singular points present in the dilaton, as is apparent from \eqref{Ricciscalar_T}.
\newline

\paragraph{$\da^2<3$ case :}

The metric has two curvature singularities at $p=0$ and $p_\eta$ (see above) and a single event horizon at $p=2$ ($\mathcal R$ is regular, $g_{tt}$ and $g_{pp}$ become respectively spacelike and timelike when crossing $p=2$). It is perfectly regular for $p\to\infty$ where matter drops out at  asymptotic infinity  (\ref{metric_asymptotic:1}). Performing the change of coordinates $q=p^{\frac1{3-\da^2}}$, the metric is asymptotically adS (remember $\Lambda<0$) iff $\da=0$. Then we need $\ga\to\infty$ to preserve $\ga\da=1$ : the dilaton potential becomes trivial (pure comological constant) and the Maxwell term in the action cancels out. Quite logically, upon imposing $\da=0$, we obtain planar Schwarzschild-adS :
\beq
	\ud s^2 =-\l(q^2-\frac2q\r)\ud t^2 -\frac{3\ud q^2}{\Lambda(q^2-\frac2q)} +q^2\l(\ud z^2+\ud\varphi^2\r)
\eeq
with $q=p^{\frac13}$.
\newline

\paragraph{$\da^2>3$ case :}

Here, we need to start the whole procedure again. Indeed, the rescaling $ \bar p = \frac{3-\da^2}{|\la|}p -\frac h{|\la|}+1$ is not valid anylonger as it changes the nature of the coodinate $p$. The same analysis but with $ \bar p = \frac{\da^2-3}{|\la|}p +\frac h{|\la|}+1$ yields the same expressions for the metric, the dilaton and the Maxwell field with changing $\eta$ to $-\eta$. Space-time is singular when $p\to\infty$ and regular at $p=0$, so this brings us to consider the change of coordinate $p=\frac1q$. As a consequence, we respectively have a singularity at $q=0$ and $q_\eta=\frac1{p_{-\eta}}$, a horizon at $q=\half$ which hides both singularities (we always get $q_\eta<\half$) and asymptotic infinity at $q\rightarrow \infty$.
Staticity inside the "de Sitter"-like horizon requires $\Lambda>0$ (the coordinate system is now valid for $0<q<\half$). The solution has a naked singularity with a cosmological horizon. We can never reduce the solution to de Sitter since this could happen only for $\da=0$, which is outside the range considered. This also explains why we cannot get Schwarzschild-dS as a solution. However when $\Lambda<0$ we have a black hole solution in the range $q>\half$ which again is never asymptotically adS given $\delta^2>3$.
\newline

\paragraph{$\eta=0$ case : }

For the sake of simplicity, we carry out the analysis for $\eta=0$ only for the regular electric solutions. The curvature singularity at $p_\eta$ drops out and we are left with a black hole solution with a single curvature singularity for $p=0$ (or $p\to\infty$), asymptotic infinity at $p\to\infty$ (or $p=0$) and an event horizon for $p=2$. The expression for the Ricci scalar simplifies to,
\beq
	\mathcal{R} = \frac{4\Lambda e^{-\da\Phi_0}}{(1+\da^2)(\da^2-3)^2}\l[\l(3\da^6-5\da^4-3\da^2-3 \r)p + 4\da^2(1-\da^2)^2 \r]p^{\frac{3\da^4-2\da^2+3}{(1+\da^2)(\da^2-3)}}
	\label{Ricciscalar:3}
\eeq
It will be singular at $p=0$ and regular when $p\to\infty$ iff
\bsea
	\frac{3\da^4-2\da^2+3}{(1+\da^2)(\da^2-3)}<0 &\Longrightarrow & \da^2<3 \nn\\
	\frac{4\da^2(\da^2-1)}{(1+\da^2)(\da^2-3)}<0  &\Longrightarrow & 1<\da^2<3 \nn
\esea
So, in the range $\da^2<1$, the solution is plagued by two curvature singularities, both at $p=0$ and $p\to\infty$; for $\da^2>3$, we can apply the same trick as above to get a cosmological solution with an initial singularity and a cosmological horizon and a black hole solution for $\Lambda<0$. The other scalar invariants show the same behaviour. Asymptotically, and performing the change of coordinates $q=p^{\frac{(\da^2-1)^2}{(\da^2+1)(3-\da^2)}}$, the metric becomes :
\beq
	\ud s^2 = -q^{2\frac{1+4\da^2-\da^4}{(1-\da^2)^2}}\ud t^2 -\frac{e^{\da\Phi_0}}{\Lambda(3-\da^2)}q^{2\frac{1+\da^2}{(\da^2-1)}}\ud q^2 +q^2(\ud z^2+\ud\varphi^2),
	\label{solution:3}
\eeq
which is asymptotically adS only if $\da=0$, just as previously. Then, the Ricci scalar smoothly goes to $\mathcal R=4\Lambda$ which is what is expected for the Einstein plus cosmological constant theory.
\newline

\paragraph{Magnetic $\eps=1$ solutions :}

Setting $\eps=1$ in (\ref{metric:1}) and $h$ as for the black hole solutions above, we obtain a "solitonic" version of \eqref{solution:2}, with Wick rotated $t=i\theta$ and $\varphi=i\tau$. This solution  is of axial symmetry at the origin $p=2$ and has a conical singularity given by $(g_{\theta\theta})'$ evaluated at $p=2$. The conical singularity can be removed by adequately rescaling the $\theta$ angle's periodicity in the standard way given that we have infinite proper distance in $p$. Whenever there is a conical singularity, metric (\ref{solution:2}) describes the gravitational field of a magnetic straight cosmic string immersed in the $(\tau, z)$ plane.
\newline

\paragraph{Magnetic dual solutions : }
As we were careful to write every quantity wrt $\da$, in order to obtain the dual magnetic solution, we only need to replace the electric Maxwell field with its magnetic dual :
\beq
	\mathcal{A}=-Qz \,\ud\varphi ,
\eeq
Then (\ref{solution:2}) is a solution for (\ref{action}) for the theory with  $\ga\da=-1$.

\subsubsection{String case : $\ga=\da=\pm 1$}
\label{subsubsection:ka0_string}

In this case, \eqref{rel_h-c-lambda:1} imposes very severe constraints on $h$ and $c$ : $h=c=0$ and questions our gauge choice for $a$ (\ref{X:1}). We can try a different approach by setting
\beq
	\Delta_X=b^2\la^2,
\eeq
in order to relax (\ref{X:1}). Unfortunately, although the system is still fully integrable, this does not yield any black hole solutions other than for $|b|=1$ (by imposing homogeneous 2-dimensional spatial sections and regularity at $p=2$). So, setting $\da=1$ and $h=c=0$  into \eqref{metric:2}, we get the following solution :
\bsea
	\dot A(p) &=& e^{-\frac{\Phi_0}{2}}\sqrt{2(1-\eta^2)}\frac1{(\eta p +1 -\eta)^2}, \slabel{A:4}\\
	e^{\Phi} &=& e^{\Phi_0}(\eta p+1-\eta),\slabel{Phi:4} \\
	\ud s^2 &=& -\frac{p(p-2)}{\eta p+1-\eta} \ud t^2 -\frac{e^{\Phi_0}}{2\Lambda}\frac{\eta p+1-\eta}{p(p-2)}\ud p^2 +(\eta p+1-\eta)\big( \ud z^2 + \ud \varphi^2 \big), \slabel{metric:4}
	\label{solution:4}
\esea
and Ricci scalar :
\bea
	\mathcal{T}_1 &=& -\eta^2\Lambda e^{-\da\Phi_0}\frac{p(p-2)}{\eta p-\eta p_\eta}   \nn \\
	\mathcal{T}_2 &=&-\frac{4\Lambda e^{-\Phi_0}}{\eta p -\eta p_\eta} \nn \\
	\mathcal R &= & \Lambda e^{-\Phi_0}\frac{3\eta^2p^2-2\eta(3\eta-4)p+4(\eta-1)^2}{\l(\eta p-\eta p_\eta\r)^3}
	\label{Ricciscalar:4}
\eea
The only singularity is at $p=p_\eta$ which for $0<\eta<1$ is covered by 2 horizons, at $p=0$ and $p=2$.  Setting $r^2=p-p_\eta$ and coordinate transforming (\ref{solution:4}) makes this obvious. This solution has therefore the same horizon structure, but not the same asymptotics,  as a planar Reissner-Nordstr\"om black hole embedded in adS. Setting $\eta=0$ effectively dimensionally reduces the geometry. 
As we will see once we get to the study of non-planar black holes, it is the endpoint in phase space of the $\ka\neq0$, $\ga\da=1$ spherically symmetric  black holes \eqref{CHMi} reported in \cite{CHM}, confirming the $\ka=0$ subspace to be the boundary of $\ka\neq0$ solutions as stated in \cite{wiltshire}. Indeed, these are, with the equivalent $\ka\neq0$ solutions \eqref{CHMi}, the only black hole solutions where the central singularity is screened by two event horizons. 

Lastly we note that it is easy to see that a regular instanton can be constructed for the case $\eta=0$  since then the temperatures of both horizons are equal in magnitude an thus the nodal singularity can be removed.  We get,
\bsea
	\dot A(p) &=& e^{-\frac{\Phi_0}{2}}\sqrt{2},\\
	e^{\Phi} &=& e^{\Phi_0}, \\
	\ud s^2 &=& -p(p-2) \ud t^2 -\frac{e^{\Phi_0}}{2\Lambda}\frac{1}{p(p-2)}\ud p^2 +\big( \ud z^2 + \ud \varphi^2 \big).
\esea

\subsubsection{$\da^2=3$ solutions}
	\label{subsubsection:ka0_gen_da_sq3}

Let us write directly the general electric solution here (as before, $\eps=1$ solutions are obtained by exchanging $t-$ and $\varphi -$ coordinates):
\bsea
	 A(p) &= & \frac{3a}{2Q}+\frac{3|\la|}{2Q}\frac{\l[\eta(p-p_0) \r]^{\frac1h}-1}{\l[\eta(p-p_0) \r]^{\frac1h}+1} \slabel{A:5}\\
	e^{\Phi} &=& e^{\sqrt3\Phi_0}\l\{\l[\eta(p-p_0) \r]^{\frac1h}+1\r\}^{\frac32}e^{-\frac{\sqrt3p}h}\l(p-p_0\r)^{-\frac3{4h}+\frac9{8h^2}-\frac{3\sqrt6c}{4h}+\frac{9c^2}{4h^2}},		 
	\quad e^{\sqrt3\Phi_0}=\l[-\frac{Q^2}{3\la^2\eta^{\frac1h}}\r]^{\frac34}\al_0^3e^{\frac{3\sqrt6}{4}\Psi_0}\slabel{Phi:5}\\
	\ud s^2 &=&-\al_0^4e^{-\sqrt3\Phi_0+\sqrt6\Psi_0-\frac{p}h}\l\{\l[\eta(p-p_0) \r]^{\frac1h}+1\r\}^{-\frac32}\l(p-p_0\r)^{\frac3{4h}+\frac3{8h^2}-\frac{\sqrt6c}{4h}+\frac{3c^2}{4h^2}}	\ud t^2 \nn \\
	&&+\frac{e^{\sqrt3\Phi_0-\frac{3p}{h}}}{2h\Lambda}\l\{\l[\eta(p-p_0) \r]^{\frac1h}+1\r\}^{\frac32}\l(p-p_0\r)^{-1-\frac3{4h}+\frac9{8h^2}-\frac{3\sqrt6c}{4h}+\frac{9c^2}{4h^2}}\ud p^2 \nn \\	
	&&+\frac{ h\la^2e^{\sqrt3\Phi_0-\frac ph}}{2\al_0^2\Lambda}\l\{\l[\eta(p-p_0) \r]^{\frac1h}+1\r\}^{\frac32}\l(p-p_0\r)^{1-\frac3{4h}+\frac3{8h^2}-\frac{3\sqrt6c}{4h}+\frac{3c^2}{4h^2}}\ud z^2 \nn \\	 
	&&+ \al_0^{-2}e^{\sqrt3\Phi_0-\sqrt6\Psi_0-\frac ph}\l\{\l[\eta(p-p_0) \r]^{\frac1h}+1\r\}^{\frac32}\l(p-p_0\r)^{-\frac3{4h}+\frac3{8h^2}+\frac{\sqrt6c}{4h}+\frac{3c^2}{4h^2}}\ud \varphi^2\slabel{metric:5} \\
	p_0&=&-\frac{3c^2}{4h}-\frac3{8h}\nn
	\label{solution:5}
\esea
From \eqref{Phi:5}, we can deduce that $h$ has to be of the form $\frac1{2n+1}$ with $n$ an integer and $\eta<0$, otherwise the sign of $e^{\Phi_0}$ is not well-defined. Also, examination of the $pp-$ and $zz-$ metric elements tells that
\beq
	h\Lambda>0.
\eeq
This tells us that the sign of $\Lambda$ will determine the sign of $h$, and vice-versa. Let us specialize to black hole solutions, by the same procedure used in the previous subsections to regularize the horizon and the singularity at $p_0$ :
\beq
	h=\sqrt6 c=1 \Rightarrow \Lambda>0, \; p_0=-\half.
\eeq
Then, rescaling some of the overall factors and taking $p\rightarrow -p$, we get
\bsea
	 A(p) &= & \sqrt{\frac{-3\eta}4}\l[a+\frac{p+p_\eta-1}{p-p_\eta}\r] \slabel{A:6}\\
	e^{\Phi} &=& e^{\sqrt3\Phi_0}\l[-\eta(p-p_\eta) \r]^{\frac32}e^{\sqrt3p} \slabel{Phi:6}\\
	\ud s^2 &=&e^{p}\l[-\eta(p-p_\eta) \r]^{-\frac32}\l(p-\half\r)\ud t^2 -\frac{e^{\sqrt3\Phi_0+3p}}{2\Lambda\l(p-\half\r)}\l[-\eta(p-p_\eta) \r]^{\frac32}\ud p^2 \nn \\	
	&&+e^{p}\l[-\eta(p-p_\eta) \r]^{\frac32}\l(\ud z^2+\ud\varphi^2\r) \slabel{metric:6}.
	\label{solution:6}
\esea
Thus, we have $p_\eta=\half+\frac1\eta<\half$ and computation of the Ricci scalar gives
\bea
	\mathcal{T}_1&=& 3\Lambda(\eta p-1)^2(p-\half)e^{-\sqrt3\Phi_0-3p}\l[-\eta(p-p_\eta) \r]^{-\frac72} \nn \\
	\mathcal{T}_2 &=& -4\Lambda e^{-\sqrt3\Phi_0-3p}\l[-\eta(p-p_\eta) \r]^{-\frac32} \nn \\
	\mathcal R &=& \mathcal{T}_1+\mathcal{T}_2 = P_3(p,\eta)e^{-\sqrt3\Phi_0-3p}\l[-\eta(p-p_\eta) \r]^{-\frac72}.
\label{Ricciscalar:6}
\eea
This solution displays two curvature singularities, at $p\to-\infty$ and $p_\eta=\half+\frac1\eta<\half$ for all $\eta<0$, as is required by \eqref{Phi:5}. We thus have $-\infty<p_\eta<\half$, depending on the value of $\eta$, but both singularities are always screened by a event horizon at $p=\half$. Asymptotic infinity is regular, but the coordinate $p$ is timelike when $p>\half$ and spacelike when $0<p<\half$. The solution is therefore cosmological.

Let us now look at the $h=-1$ solutions, which have $\Lambda<0$. We get :
\bsea
	 A(p) &= &  \sqrt{\frac{-3\eta}4}\l[a+\frac{\eta(p-\half)-1}{\eta(p-\half)+1}\r] \slabel{A:6bis}\\
	e^{\Phi} &=& e^{\sqrt3\Phi_0}\l[\eta(p-\half)+1 \r]^{\frac32}e^{\sqrt3p}, \quad e^{\sqrt3\Phi_0}=\l[-\frac{Q^2}{3\la^2\eta}\r]^{\frac34}\al_0^3e^{\frac{3\sqrt6}{4}\Psi_0} \slabel{Phi:6bis}\\
	\ud s^2 &=&-e^{p}\l[\eta(p-\half)+1 \r]^{-\frac32}\l(p-\half\r)\ud t^2 -\frac{e^{\sqrt3\Phi_0+3p}}{2\Lambda\l(p-\half\r)}\l[\eta(p-\half)+1 \r]^{\frac32}\ud p^2 \nn \\	
	&&+e^{p}\l[\eta(p-\half)+1 \r]^{\frac32}\l(\ud z^2+\ud\varphi^2\r) \slabel{metric:6bis}.
	\label{solution:6bis}
\esea
We still need $\eta<0$, but now
\bea
	\mathcal R &\sim&-4\Lambda e^{-\sqrt3\Phi_0-3p}\l[1+\eta(p-\half) \r]^{-\frac32}.
\label{Ricciscalar:6bis}
\eea
so there is a curvature singularity both at $p\to-\infty$ and $p_\eta=\half-\frac1\eta$. $p=\half$ and $p\to+\infty$ are regular points, but $p=\half$ does not screen anylonger the $\eta$-singularity.

\subsection{Solutions for arbitrary $\gamma$ and $\delta$}
	\label{subsubsection:ka0_arbitrary}

When seeking a solution of \eqref{radial_eq} for general coupling constants $\ga$ and $\da$ one has to make some suitable ansatz. Given the form of the general solution, found in the previous section, we expect by continuity some form of polynomial solutions in the same coordinate system. As we saw earlier on, roots of the polynomial are singular points, either of the coordinate system or of the spacetime metric. The generic ansatz that works for $B(p)$ for arbitrary coupling constants is a second-order polynomial in $p$. Setting
\beq
	u=\ga^2-\ga\da+2, \quad v=\da^2-\ga\da-2, \quad w=\frac 1u\Big[s(3-\da^2)+(1-\ga\da)^2)\Big],
\eeq
\beq
	\la^2=\Delta_X = \Big(\frac{sh}u\Big)^2+\frac{wu}{s(v+u)}\Big[2c^2+\Big(\frac{sh}u\Big)^2\Big], \quad \bar p = \frac w{|\la|}p-\bar h+1, \quad \bar h=\frac{sh}{u|\la|}, \quad \bar c=\frac{c}{|\la|},
\eeq
gives for $A(p)$ and $X(p)$ :
\bea
	A(p) &=& a-\frac{1-\ga\da}wh-\frac{sv}{wu}(p-1) \label{A:7}\\
	X(p)&+&\frac{Q\eps}2(1-\ga\da)B(p)=\frac {\la^2}{2w}p(p-2), \label{X:7}
\eea
where we have dropped all bars and our coordinate $p$ is dimensionless. Notice that $uw$, $u$ and the discriminant of $X$ are necessarily positive for all $\da^2<3$, and consequently so is $w$. Thus, for $\da^2<3$, the $p-$coordinate is  spacelike. If $\da^2>3$, no general arguments can easily be made. The roots of $X(p)$, namely $p=0$ and $p=2$, are again the singular points of the solution. It is easy to check that setting $\ga\da=1$ in the above gives us back the solutions with $\eta=0$ ($A(p)$ must be polynomial), cf \eqref{m7d} discovered in the previous section. However, case $\eta\neq0$ has no equivalent here. The solution we obtain for general couplings $\ga$ and $\da$ is the following :
\bsea
	\ud s^2&=&  - p^{C(  h,  c,\eps)}(p-2)^{C(-  h,-  c,\eps)}\ud t^2 +\frac{e^{\da\Phi_0}}{-w\Lambda}p^{F(  h,  c)}(p-2)^{F(-  h,-  c)}\ud p^2+ \nonumber\\
			&&+p^{B(  h,  c)}(p-2)^{B(-  h,-  c)}\ud z^2 + p^{C(  h,  c,-\eps)}(p-2)^{C(-  h,-  c,-\eps)}\ud\varphi^2, \slabel{metric:7}\\
	e^{\Phi} &=& e^{\Phi_0}p^{D(h,c)}(p-2)^{D(-h,-c)} \slabel{Dilaton:7},
\label{solution:7}
\esea
where $\Phi_0$ is an integration constant,  and the exponents are
\bea
	F(  h,  c) &=& \frac{1-  h}w\Big[\da^2+\frac{\ga\da}s(1-\ga\da)\Big]-1-\frac1s(\da\sqrt2  c+\ga\da), \\
	B(  h,  c) &=&\frac{1-  h}w\Big[\da^2-2+\frac{\ga\da}s(1-\ga\da)\Big]+1-\frac1s(\da\sqrt2  c+\ga\da), \\
	C(  h,  c,\eps) &=&\frac{1-  h}w\Big[1+\frac\eps s(1-\ga\da)\Big]-\frac\eps s(1-\ga\sqrt2  c),\\
	D(h,c) &=& -\frac\ga s+\frac{\ga+\da}{ws}(1-h)-\frac{\sqrt2c}s.
	\label{powers}
\eea
The roots $p=0$ and $p=2$ are again interchanged under the symmetry (\ref{symmetry}) and as a result the solutions $p>2$ or $p<0$ are equivalent, up to inversing the signs of both $h$ and $c$. Inversing the sign of $\eps$ allows us to get the magnetic solutions from the electric ones, so we will consider the case $\eps=-1$ in the following without any loss of generality. As before, the sign for the cosmological constant and the nature of spacetime will depend on the couplings $\gamma$ and $\delta$. Indeed, for $w>0$ we see that $\Lambda$ is negative and the coordinate $p$ is timelike in between 0 and 2, whereas for $p>2$ the solution is static.
Given the form of the metric and its symmetry in $h$ and $c$ we can find with ease the form of the metric for large $p$ by setting $h=c=0$. We obtain, given the coordinate transformation
\beq
	q=p^{\frac{(\ga-\da)^2}{wu}},
\eeq
the solution of maximal symmetry
\beq
	\ud s^2 = -q^{2\frac{\ga^2-\da^2+4}{(\ga-\da)^2}}\ud t^2 +q^2\l(\ud z^2+\ud\varphi^2\r) +\frac{wu^2e^{\da\Phi_0}}{(-\Lambda)(\ga-\da)^4}q^{2\frac{\da+\ga}{\da-\ga}}dq^2.
	\label{metric_asymptotic:7}
\eeq
This can be locally AdS if and only if we take the limit $\da=0$, $\ga\to\infty$, which is similar to the previous section, see \eqref{metric_asymptotic:1}. Here again, this amounts to cancelling the Maxwell term ($\dot A=0$ in this limit) and taking the dilaton to be trivial, so, once again, this is in agreement with Wiltshire \emph{et al.} \cite{wiltshire}.

Furthermore, the Ricci scalar goes like :
\bea
	\mathcal T_1 &=& -\frac{2\Lambda}{ws^2}e^{-\da\Phi_0}\l[(\ga+\da-\ga w)p+w(\ga+\sqrt2c)+(\ga+\da)(h-1)\r]^2\l(p-2\r)^{-1-\da D(-h,-c)}p^{-1-\da D(h,c)} \nn \\
	\mathcal T_2 &=& 4\Lambda e^{-\da\Phi_0}\l(p-2\r)^{-\da D(-h,-c)}p^{-\da D(h,c)} \nn \\
	\mathcal{R} &=&-\mathcal T_1-\mathcal T_2\sim_\infty p^{\frac{4\da(\ga-\da)}{wu}}, \label{Ricciscalar:7}
\eea
which given that $wu>0$ for $\da^2<3$, yields the regular asymptotic region for $\ga<\da$, $\da>0$ and $p$ spacelike infinity (if $\da<0$, remember there is a symmetry : change simultaneously $\ga\to-\ga$, $\da\to-\da$, $\Phi\to-\Phi$). Upon taking the limit $\da=0$, $\ga\to\infty$, the Ricci scalar equals $4\Lambda$ as expected.

The Weyl square exhibits the same kind of behaviour both locally in $p$ and $p-2$, and asymptotically. However, the exact expression is not enlightening.

\subsubsection{Polynomial black hole solutions}
	\label{subsubsection:ka0_arbitraryBH}

To get the black hole solutions, we need to set (note that this is the equivalent of $h=\da\sqrt2c$ as in the previous subsection, for arbitrary couplings)
\beq
	h=\frac{\ga+\da}{s}\sqrt2c \ ,
\eeq
which yields
\beq
  \tilde h = \frac{\ga^2-\da^2}{2u}, \quad \sqrt2\tilde c = \frac{\ga-\da}2 \ .
\eeq
Then, for the electric case $\eps=-1$,
\bsea
	e^{\Phi} &=&e^{\Phi_0}p^{\frac{4(\da-\ga)}{3\ga^2-\da^2-2\ga\da+4}}, \slabel{phi_0_pol} \slabel{Dilaton:8}\\
	A(p) &=& 2e^{-\frac{\ga\Phi_0}2}\sqrt{\frac{-v}{uw}}\l[p-1+\frac vu\l[\l(1-\ga\da\r)h-wa\r] \r],\slabel{A:8}\\
	\ud s^2 & = &(p-2)p^{\frac{-\ga^2-\da^2+2\ga\da+4}{3\ga^2-\da^2-2\ga\da+4}} \ud t^2-\frac1{w\Lambda}e^{\da\Phi_0}p^{\frac{-3\ga^2+5\da^2-2\ga\da-4}{3\ga^2-\da^2-2\ga\da+4}} \frac{\ud p^2}{p-2} \nonumber\\
			&& +p^{\frac{2(\ga-\da)^2}{3\ga^2-\da^2-2\ga\da+4}} \big(\ud z^2+\ud \varphi^2),\slabel{metric:8}
	\label{solution:8}
\esea
where we have rescaled the coordinates and reexpressed the extra integration constants in terms of $\Phi_0$.

The Ricci scalar is now :
\bea
	\mathcal T_1 &=& -\frac{8\Lambda\l(\ga-\da\r)^2}{wu^2}e^{-\da\Phi_0}\l(p-2\r)p^{-1+\frac{4\da(\ga-\da)}{wu}} \nn\\
	\mathcal T_2 &=& 4\Lambda e^{-\da\Phi_0}p^{\frac{4\da(\ga-\da)}{wu}} \nn \\
	\mathcal{R} &=& -\mathcal T_1 - \mathcal T_2 =  P_1(p)p^{-1+\frac{4\da(\ga-\da)}{wu}}, \label{Ricciscalar:8}
\eea
The exponent of $p$ is negative for all $\da^2<3$, which will be the case for the five-dimensional metrics in particular. This way, we have an event horizon at $p=2$ and a curvature singularity at $p=0$.

The $\ga=\da$ case here is trivial, which is what we expected from \eqref{solution:7}. By setting $\ga=\da=1$, such a solution can be related to the string case of \ref{subsubsection:ka0_string} for $\eta=0$, which gives flat space. However, it is possible to find a non-trivial solution, but we need to go to higher order polynomials in $p$.

It is also worth noting that this is the only solution we could obtain for $\da=0$ (pure cosmological constant) with a non-trivial dilaton. Nevertheless, its asymptotics are not regular, so it is not in contradiction with \cite{wiltshire}. Black hole solutions \eqref{solution:8} were introduced in \cite{Cai}, for special relations between the couplings $\ga$ and $\da$, whereas here, the couplings have been kept arbitrary from the start.
\newline

\paragraph{Magnetic dual solutions :}

Here again, we can use the same procedure as before and obtain a magnetic dual solution to this one. We just need to set $\ga\to-\ga$ in the previous metric and take
\beq
	\mathcal A=-Qz\ud\varphi
\eeq
as Maxwell field. This will be particularly useful since in this way we can get an extra upliftable solution when $\ga\da=-1$ (giving $\ga\da=1$ after use of the duality).

\subsubsection{Higher-order solution}
	\label{subsubsection:ka0_higher_order}

Solving for $B(p)=B_N(p-M)^N+B_2p^2+B_1p+B_0$, we find a unique non-trivial solution for $\ga=\da$, $h=c=0$ :

\bea
A(p)& =& 2(1-\da^2)(3-\da^2)B_Np^{-\frac{(1+\da^2)}{(3-da^2)}} -\frac{2a}{\eps Q(1+\da^2)}+ \frac{2(3-\da^2)M}{(1+\da^2)\eps Q}\\
Y(p) &=& \frac{(3-\da^2)}2p^2+2\frac{(3-\da^2)}{(1+\da^2)}Mp-\frac12QB_N(1-\da^2)p^{2\frac{(1-\da^2)}{(3-\da^2)}}
\eea 
where we have translated $p$ to bring the origin to $p=0$ and $M$ is arbitrary and linked to the mass of the solution, which reads :
\bsea
	e^\Phi &=& e^{\Phi_0}p^{\frac{2\da}{3-\da^2}} \slabel{Phi:9} \\
	\ud s^2 &=& -Yp^{2\frac{2-\da^2}{\da^2-3}}\ud t^2 -\frac{e^{\da\Phi_0}}{2\Lambda Y}p^{\frac{2\da^2}{3-\da^2}}\ud p^2 +p^{\frac{2}{3-\da^2}}(\ud z^2+\ud \varphi^2) \slabel{metric:9}
	\label{solution:9}
\esea
where $Q$, $B_N$ and $\Phi_0$ are not independent. This solution is not valid for $\ga=\da=\sqrt3$, so this will have to be studied separately.

The same argument as in section \ref{subsubsection:ka0_general} can be made about staticity : the solution is static iff $\Lambda<0$ and $\da^2<3$ or $\Lambda>0$ and $\da^2>3$. Let us not enter in the details once again and concentrate on the former case.

Asymptotically, this metric goes to
\beq
	\ud s^2 = p^{\frac{2}{3-\da^2}}\l(-\ud t^2+\ud z^2+\ud \varphi^2\r)+\frac{e^{\da\Phi_0}}{\l(3-\da^2\r)\l(-\Lambda\r)}p^{2\frac{2\da^2-3}{3-\da^2}}\ud p^2,
\eeq	
which becomes after the coordinate transformation $q=p^{\frac{1}{3-\da^2}}$ :
\beq
	\ud s^2 = q^2\l(-\ud t^2+\ud z^2+\ud \varphi^2\r)+\frac{\l(3-\da^2\r)e^{\da\Phi_0}}{\l(-\Lambda\r)}q^{2\da^2-2}\ud q^2,
	\label{metric_asymptotic:9}
\eeq	
and it is of course asymptotically adS iff $\da=0$. All of this is confirmed by the Ricci scalar :
\beq
	\mathcal{R}=\frac{2\Lambda}{(3-\da^2)^2}e^{-\da\Phi_0}\Big[ -3(\da^2-2)(\da^2-3)-\frac{2\da^2M}p+\eps QB_3\da^2(\da^2-1)p^{\frac{4}{\da^2-3}}\Big]p^{\frac{2\da^2}{\da^2-3}} \sim_\infty p^{\frac{2\da^2}{\da^2-3}}.
	\label{Ricciscalar:9}
\eeq
The Ricci scalar exhibits a curvature singularity at $p=0$ and there will be an event horizon where there is a zero of $Y(p)$. Once more, the correct asymptotic behaviour is for $\da^2<3$. For $\da=0$, we get the expected value for the Einstein plus cosmological constant theory.

Setting $\ga=\da=0$, the metric now reads :
\bea
	\ud s^2 &=& -V(q)\ud t^2+\frac{9\ud q^2}{2(-\Lambda)V(q)} + q^2\l(\ud z^2+\ud\varphi^2\r) \\
	V(q) &=& \frac32q^2+\frac Mq-\frac{QB_N}{2q^2},\quad q=p^{\frac13} \nn
\eea
which is just planar Reissner-Nordstrom-adS : contrarily to previous cases, taking this limit does not cancel out the Maxwell contribution.

Comparing with the string case \ref{subsubsection:ka0_string}, we find that the metric \eqref{metric:9} is the generalization of the metric \eqref{metric:4} for arbitrary $\ga=\da$, and $\eta=1$ (constant Maxwell field).
This solution was first found in \cite{Cai}.

\subsubsection{Limiting case $\ga=\da=-\sqrt3$}
	\label{subsubection:ka0_gasqrt3_dasqrt3}

The master equation \eqref{radial_eq} can be solved directly for certain values of the coupling constants. For example taking $\ga=\da=-\sqrt3$ we have,
\beq
	\eps Q\ddot{\overbrace{\log B}}+2\dot{\l(\frac{p}{B}\r)}+\frac{k}{B^2}-a\frac{\dot B}{B}-\frac{\ddot B}{B}\l[-\l(\frac a2+h\r)p-\frac{c^2}4-\frac{a^2}8-\eps\frac{Qk}2B\r].
		\label{radial_eq:10}
\eeq
Considering $k=c=h=a=0$, rescaling  $\bar p=\frac{\mu p}{\eps Q}$ and then dropping the bars yields the solution :
\bea
	B(p)&=&\lambda e^{-p}-\frac{2\eps Q}{\mu^2}(p-1), \label{B:10}\\
	Y(p)&=&X(p)-\eps QB(p)=-\eps Q\lambda e^{-p}+\frac{2Q^2}{\mu^2}(p-1). \label{X:10}
\eea
Here, $\mu,\la$ are integration constants.
The resolution is then straightforward and the solution reads, for $\eps=-1$ :
\bsea
	A(p)&=&\l(\lambda e^{-p}-\frac{2Q}{\lambda\mu^2}\r)\sqrt{\lambda Q}e^{\frac{\sqrt3\Phi_0}2}, \slabel{A::10} \\
	e^{\Phi}&=&e^{\Phi_0}e^{-\frac{\sqrt3}2p}, \slabel{Dilaton::10} \\
	\ud s^2&=&-Y(p)e^{\frac p2}\ud t^2-\frac{Q^2e^{-\sqrt3\Phi_0}}{2\Lambda\mu^2Y(p)}e^{\frac32p}\ud p^2+e^{\frac p2}\l(\ud z^2+\ud\varphi^2\r)\slabel{metric::10}.
	\label{solution:10}
\esea
We have set $\eps=-1$ in the metric expression in order to get a black hole.  Staticity outside the hole imposes $\Lambda<0$, so we have an adS-like solution. The solution verifies $\lambda Q>0$ and admits several horizons. To check this, let us calculate the Ricci scalar,
\beq
	\mathcal R = -\frac{\Lambda}{4Q^2}e^{\sqrt3\Phi_0}\l(3\lambda\mu^2 e^{-p}+6Q^2p-22Q^2\r)e^{-\frac32p}
\eeq
which is regular for all possible finite zeros of $X(p)$, singular as $p\to-\infty$ and cancels as $p\to\infty$. The same behaviour is exhibited by the Weyl square. There is a curvature singularity as $p\to-\infty$  screened by two horizons iff
\beq
	\la Q<2\frac{Q^2}{\mu^2},
\eeq
the inequality being saturated in the extremal case. Again the horizon structure is similar to planar RN in adS. Finally, to keep our radial coordinate spacelike as we approach asymptotic infinity, we need
\beq
\frac{\mu}{Q}<0 \textrm{ if } \eps=-1.
\eeq

\section{Maximally symmetric solutions in 4-dimensional spacetime}

We start this section by very briefly considering the case $\Lambda=0$ which yields insight on our case of interest $\Lambda\neq 0$. This case was first analysed Gibbons and Maeda \cite{gibbonsmaeda} and later on revisited in the case of $\ga=\da=1$ by Horowitz et al \cite{horowitz}. In the coordinate system (\ref{X}), it is trivial to integrate since from (\ref{m1})
\beq
\beta(x)\equiv \alpha^2=\kappa x^2+\beta_1 x+\beta_0
\eeq
where $\beta_1, \beta_0$ are arbitrary constants. Since in that case the coupling $\da$ can be chosen at will, we fix it to be  $\da=1/\ga$ and then (\ref{int1}) is simply an identity, whereas (\ref{int2}) gives $A$ by direct integration as in (\ref{cond}). The important thing to note is that the second order coefficient of $\beta$ is directly given by $\kappa$. Whenever this is the highest order coefficient of $\beta$ this immediately means that $\Lambda=0$. According to the prescription we described in the second section we find with ease the remaining metric components obtaining the general solution for $\Lambda=0$.

Let us now consider $\Lambda\neq 0$. We have to simultaneously solve for two coupled equations (\ref{int1}) and (\ref{int2}). For $d=4$ these read :
\beq
\label{int14}
\beta \ddot{A}+\dot{\beta}\dot{A}-(\ga\da-1)(\ka x-\half \dot{\beta})\dot{A}+(a-\frac{sQ A}{2})\dot{A}=0,
\eeq
\bea
\label{int24}
&-\frac{\da^2+1}{2}\left[\frac{1}{2}\dot{\beta}-\kappa x+\frac{h}{\da^2+1} \right]^2-\kappa \beta -\half (\ddot{\beta}\beta-\dot{\beta^2})&\nn\\
	&=&\nn\\
	&\frac{h^2 (1-\ga \da)^2}{2(\da^2+1)(\ga+\da)^2}-\frac{(1-\ga\da)}{2s}\left(a-\frac{sQ}2A\right)(\dot\ba-2\ka x)+\frac1{2s}\left(a-\frac{sQ}2A\right)^2-\half Q\dot A\ba.&
	\label{al_x_max0}
\eea
It is useful to note that the coordinate systems are related via $\dot{\beta}=2p$, and therefore are the same only if $\beta$ is a second order polynomial.

As noted previously for $\kappa=0$, the case $\ga\da=1$ is special since (\ref{int14}), (\ref{int24}) decouple and furthermore $(\ref{int14})$ is integrable. For this case :
\beq
	\ba\dot A = \frac{sQ}4A^2-aA+k,
	\label{k_x_max}
\eeq
\beq
-\frac{1}{2(\da^2+1)}\left[\frac{\da^2+1}{2}\dot{\beta}-\kappa x (\delta^2+1)+h \right]^2=\kappa \beta +\half (\ddot{\beta}\beta-\dot{\beta^2})-\frac{Qk}{2} +\frac{a^2}{2s}.
	\label{al_x_max}
\eeq
The general solution to this equation can be obtained by numerical integration. Some explicit solutions can be obtained upon supposing that $\beta$ is of polynomial form. One of them is the $\Lambda=0$ solution discussed above and the second is a black hole solution first obtained in \cite{CHM} for $\ka=1$. The potential reads,
\beq
\beta(x)  =  \ka\frac{\da^2+1}{\da^2-1} x^2 -\frac{2h}{\da^2-1}x +(\da^2-1)\frac{Qk}{4\da^2\ka} +\frac{h^2}{\ka(\da^4-1)}-\frac{a^2(\da^2-1)}{4\ka(\da^2+1)}.
\eeq
The solution is not valid for $\ga=\da=1$. After a translation and some redefinitions of parameters the solution takes the form of \cite{CHM},
\beq
	\ud s^2 = -U(x)\ud t^2 + \frac{\ud x^2}{U(x)} +R^2(x)d\Omega^2, \qquad U=\frac{\beta}{R^2},
	\label{CHMmetric}
\eeq
where a suitable change of the origin and rescaling of constants gives
\bsea
	\beta(x) & = & \ka\frac{1+\ga^2}{1-\ga^2}x^2-2(1+\ga^2)Mx+\frac{sQ^2}{4}\,e^{-\frac{\Phi_0}{\da}}, \\
	e^{\Phi} & = & e^{\Phi_0}x^{\frac{2\da}{1+\da^2}}, \\
	\dot A(x) &= & Q x^{-2}e^{-\frac{\Phi_0}{\da}}, \\
	R^2(x)&=&x^{\frac{2\da^2}{1+\da^2}}, \\
	\Lambda & = & \frac{\ka}{1-\da^2}\,e^{\da\Phi_0},\\
	\ga\da&=&1.
	\label{CHMi}
\esea
Note the absence of an extra parameter presented in \cite{CHM} (see also \cite{Cai} for $\kappa=-1$) which can be gauged away. This solution is clearly valid only for $\ka\neq0$. The $\kappa=0$ black holes are the ones presented in the previous section. The solution has one singularity in $x=0$ and two horizons at the two roots of $\ba(x)$. The appearance of an extra horizon, compared with the case when $\ka=0$, is linked with the non-zero curvature of the horizon  ($\ka\neq0$).
The dual magnetic solution is readily obtained from \eqref{CHMi}. Using the dual potential \eqref{dualpotential} and the duality map \eqref{dualitymap}, we get the magnetic solution by simply replacing the Maxwell field of \eqref{CHMi} :
\beq
	\mathcal A = \frac{Q}{\ka}\,\textrm{co}(\theta)\,\ud\varphi ,
\eeq
and setting $\da=-\frac1\ga$ in the solution (\ref{CHMi}).

This particular solution is not defined for $\ga=\da=\pm1$. If we do try to find a solution for the string case, the only permitted polynomial solution  is one of second degree verifying :
\beq
	\kappa(\ba_2-\kappa)=0,
\eeq
where $\ba_2$ is the highest order coefficient.
Therefore we either have a toroidal black hole [cf \ref{subsubsection:ka0_string}] or a $\Lambda = 0$ solution (see \cite{gibbonsmaeda} or \cite{horowitz}). In a moment we will see that use of the duality can give magnetic string solutions.

If $\ga\da\neq1$, we have to make some starting assumption in order to solve for  $A(x)$. A simple starting point is to assume that $A$ is a linear function and from (\ref{int14}), we get :
\beq
	sQ\ddot A = 0 = (1+\ga\da)\beta^{(3)}.
	\label{A_x_max2}
\eeq
This last equation gives us two constraints: either $\beta(x)$ is a second-order polynomial or $\ga\da=-1$. Suppose then that  $\ga\da\neq \pm 1$.
Solving then for a second order polynomial in (\ref{al_x_max}) gives us three distinct possibilities. First of all $\Lambda=0$ solutions \cite{gibbonsmaeda}, or again a subclass of $\Lambda RN$ where the dilaton is trivial. The third case lies within the interest of our study and the action parameters are related via $\ga+\da=0$. The solution reads :
\bsea
	\beta(x) & = &\ba_2x^2-2\frac{(1+\da^2)}{\da^2}Mx,  \\
	e^{\Phi} & = & e^{\Phi_0}x^{\frac{2\da}{1+\da^2}}, \quad e^{\da\Phi_0} = \frac{2\Big[(1-\da^2)\ba_2+\ka(1+\da^2)\Big]}{Q^2(1+\da^2)},\\
	\dot A(x) &= & \frac{2}{(1+\ga^2)Q}\Big[(1-\ga^2)\ba_2 +\ka(1+\ga^2) \Big], \\
	R^2(x)&=& x^{\frac{2\da^2}{1+\da^2}}, \\
	\Lambda & = & \frac{\ka-\ba_2}{2}e^{\da\Phi_0},\\
	\ga+\da&=&0.
	\label{CHMii}
\esea
This is again the solution presented in \cite{CHM} and \cite{Cai}. For $\kappa=1$ it has one singularity in $x=0$ and one horizon at $x_h = \frac{2sM}{\da^2}$. In order to have the $xx$ metric element spacelike outside the horizon, we also need $\ba_2>0$.

As we noticed from (\ref{A_x_max2}) when $\ga \da=-1$ we can have a higher order polynomial. Upon making this assumption for $\beta$,
\beq
\beta(x) = \ba_N x^N+\ba_2x^2+\ba_1x+\ba_0,
\eeq
where $N$ is assumed to be different from $2$, $1$ or $0$,
we obtain :
\beq
\al_2=\ka \quad \mathrm{or} \quad \da^2=\frac13,
\eeq
which both lead to the black hole solution of \cite{CHM}
\bsea
	\beta(x) &=& \ba_Nx^{\frac4{1+\da^2}} +\ka x^2-2(1+\da^2)Mx\\
	\dot A(x) & =& \frac{4\ka}{Qs}, \\
	e^{\Phi(x)} & =& e^{\Phi_0}x^{\frac{2\da}{1+\da^2}}, \quad e^{\frac{\Phi_0}\da}=\frac{4\ka}{Q^2(1+\ga^2)},\\
	R(x)^2 &=& x^{\frac2{1+\da^2}},\\
	\Lambda&=&-\ba_N\frac{(3-\da^2)e^{\da\Phi_0}}{(1+\da^2)^2}, \\
	\ga\da&=&-1.
	\label{CHMiii}
\esea
Let us focus on the case $\delta=1$.  It actually coincides with the previous solution \eqref{CHMii} for which $\ga+\da=0$ as can be easily checked. Setting $x=r^2$ the solution reads
\beq
\ud s^2 = -r^2[(\ba_N+\kappa)r^2-\frac{4M}{r^2}]\ud t^2 + \frac{4\ud r^2}{(\ba_N+\kappa)r^2-\frac{4M}{r^2}} +r^2d\Omega^2, 
\eeq
with $e^{\Phi(r)} = e^{\Phi_0}r^{2}$. This solution is singular at $r=0$ and has an event horizon at $r_h=\sqrt{\frac{4M}{\beta_N+\kappa}}$. By use of the duality the above metric is a magnetic solution with
\beq
	\mathcal A = \frac{Q}{\ka}\,\textrm{co}(\theta)\,\ud\varphi ,
\eeq
and $\ga=1$. This is the unique $\kappa=1$ solution for the couplings $\ga=\da=1$ we could find. This is the equivalent for non-zero cosmological constant of the solution presented by \cite{horowitz}, but without the extra-singularity present in that particular solution.

For the adequate couplings we will uplift the magnetic version of \eqref{CHMiii} in order to obtain a 5-dimensional metric.

\section{Uplifted 5-dimensional solutions}

So far, we have shown how to obtain exact solutions for Einstein-Maxwell-dilaton
theories with a Liouville potential. In this section, we uplift
 $d$-dimensional Einstein-Maxwell-Dilaton solutions to $(d+1)$-dimensional Einstein solutions with a cosmological constant. Indeed the uplifted theory is just
\beq
  S= \int d^{(d+1)} x \sqrt{g^{(d+1)}} \left[ R^{(d+1)} -2 \Lambda \right]
\eeq
where $g^{(d+1)}$, $R^{(d+1)}$ and $\Lambda$ are the determinant of the
$(d+1)$-dimensional metric, the $(d+1)$-dimensional
scalar curvature and the cosmological constant, respectively.
The argument is a standard one: taking the metric ansatz
\beq
\ud s^2_{d+1}=e^{-\da\Phi}\ud s^2_{d} + e^{(d-2)\da\Phi}\l(\ud w + A_\nu\ud x^\nu\r)^2 ,
\eeq
one can reduce the $(d+1)$-dimensional theories
 to the $d$-dimensional Einstein-Maxwell-Dilaton action (\ref{action}),
where we have the relations (\ref{KK}), $\gamma\delta = 2/(d-2)$.
As an illustration, let us uplift a 4-dimensional metric to 5 dimensions.
This can be done  only
in the two cases $\ga=\pm\sqrt{3}$ and $\da=\pm\frac1{\sqrt3}$,
which satisfies the relation $\ga\da=1$.
The general way to uplift the metric is to use the relation
\beq
	\ud s^2_{5D}=e^{\mp\frac\Phi{\sqrt3}}\ud s^2_{4D}
      + e^{\pm2\frac\Phi{\sqrt3}}\l(\ud w + A_\nu\ud x^\nu\r)^2 \ ,
\eeq
where the 4-dimensional metric $\ud s^2_{4D}$ can be obtained
from the results in the previous sections.
There are two ways to get 5-dimensional solutions from 4-dimensional ones.
One is to use the electric solutions and the other is
to use the magnetic solutions. Let us discuss each case separately.

\subsection{From Electric Solutions}

Let us use the electric solutions
with $\gamma \delta =1$ to obtain the solutions of the form
\beq
 \ud s^2_{5D}=e^{-\da\Phi}\ud s^2_{4D} + e^{2\da\Phi}\l(\ud w + A_t (p) \ud t \r)^2
\eeq
where $p$ is the radial coordinate in 4 dimensions.
This gives stationary solutions in 5 dimensions.
For instance, taking as a starting point the black hole solution \eqref{solution:2}, we can uplift it to
\begin{eqnarray}
ds^2_{5D} &=& -e^{-\Phi_0/\sqrt{3}} \frac{\sqrt{p}(p-2)}{\eta p +1 -\eta} dt^2
   +\frac{3}{-8\Lambda}\frac{dp^2}{p(p-2)}
   +e^{-\Phi_0/\sqrt{3}} \sqrt{p} \left( dz^2 + d\varphi^2 \right) \nonumber\\
 &&  + e^{2\Phi_0/\sqrt{3}} \frac{(\eta p +1 -\eta)}{\sqrt{p}}
   \left[ dw + \left( a +e^{-\frac{\sqrt3\Phi_0}2}\frac1{\sqrt{1-\eta^2}} \frac{p-1+\eta}{\eta p -\eta +1 } \right)dt  \right]^2.
\end{eqnarray}
The above solution admits rotation due to the lower-dimensional electric field and has a curvature singularity at $p=0$ and an event horizon at $p=2$ and possibly also at $p_\eta=1-\frac1\eta$. Due to this, it is asymptotically only locally adS except if we set $a=-\frac1\eta e^{-\frac{\sqrt3\Phi_0}2}\frac1{\sqrt{1-\eta^2}} $.  Upon doing so we obtain for $p=r^4$ an adS patch in Poincar\'e coordinates. The static limit of the hole becomes clearly identifiable at $p_s=1+\frac1\eta>2$ if $0<\eta<1$. Then it is outside the event horizon. However, we have $p_\eta<0$ and there is  only one event horizon at $p=2$. If on the contrary $-1<\eta<0$, the static limit is outside the range of coordinates but an outer horizon appears at $p_\eta$.

Another example is to uplift the solution \eqref{CHMi}.
The resultant metric is
\bea
	\ud s^2_{5D}&=& e^{-\frac{\Phi_0}{\sqrt3}}(2\ka x+8M)\ud t^2 + e^{-\frac{\Phi_0}{\sqrt3}}\frac{\ud x^2}{-2\ka x^2-8Mx+Q^2e^{-\sqrt3\Phi_0}} \nonumber \\
	&& +e^{-\frac{\Phi_0}{\sqrt3}}\Big(\ud\theta^2+\textrm{si}^2(\theta)\ud\varphi^2\Big)
      +e^{\frac{2\Phi_0}{\sqrt3}}x\,
                 \Big(\ud w-\frac{2Q}xe^{-\sqrt3\Phi_0}\ud t\Big)^2 .
\eea
From the computation of Kretschmann invariant, it turns out the above metric is
  perfectly regular. This can be explained in the light of \cite{GHT}.
 Indeed, it can be shown that, following compactification, the appearance of a dilaton is accompanied by singularities. Uplifting 4-dimensional solutions to
  higher-dimensional ones will then smooth out those singularities,
  which are just an artifact of the compactification to 4 dimensions in our case.
  More generally, all uplifted solutions here displayed are less singular than in four dimensions.

\subsection{From Magnetic Solutions}

From here on, we uplift the 4-dimensional magnetic solutions with $\gamma \delta =1$
to 5 dimensions. In order to do this, we simply 
 apply the duality transformation to the electric solutions with
$\gamma \delta =-1$ to get the magnetic solutions with $\gamma \delta =1$.
Once we obtain the magnetic solutions with $\gamma \delta =1$, we can uplift
the solutions to 5 dimensions according to the metric anzatz
\beq
 \ud s^2_{5D}=e^{-\da\Phi}\ud s^2_{4D} + e^{2\da\Phi}\l(\ud w + \omega  \ud \varphi \r)^2,
\eeq
where in 4 dimensions $\omega$ is a function of the spatial coordinate $z$ ($\kappa =0$) or $\theta$ ($\kappa =\pm 1$).
These methods can be applied both to the cylindrically symmetric spacetimes
and maximally symmetric cases.
In this way, we obtain black holes with non-trivial horizons in 5 dimensions.
Let us present several examples.

\subsubsection{Nil horizon}

Using the duality \eqref{dualpotential} and \eqref{dualitymap}, we can map a 4-dimensional electric metric with $\ga\da=-1$ to a 4-dimensional magnetic metric with $\ga\da=1$.
By the duality transformation, the 4-dimensional metric and the dilaton do not change.
If $\kappa =0$, the duality relation \eqref{dualpotential}
yields
\bea
   \partial_z \omega (z) = -e^\Omega \alpha^{-1} A' = -Q \ .
\eea
Hence, we have the vector potential
\beq
	\mathcal A = -Q z\ud\varphi
\eeq
through the duality transformation.
Using equations \eqref{Dilaton:8} as well as \eqref{metric:8},
and setting $\ga=\sqrt3$ and $\da=-\frac1{\sqrt{3}}$ (so that $\ga\da=-1$),
we get the following
\bea
	\ud s^2_{5Dmag}&=&-(p-2)p^{-\frac5{11}}\ud t^2
      -\frac{9}{22\Lambda}\frac{\ud p^2}{p(p-2)} \nonumber \\
&&+p^{\frac4{11}}\l(\ud z^2 + \ud \varphi^2 \r)
 +p^{\frac8{11}}\Big(\ud w^2 - Q z \ud\varphi \Big)^2 \ .
	\label{metric5d_0_mag}
\eea
The square of the Weyl tensor of the above solution can be  calculated as
\beq
	C = \frac{2\Lambda^2(1603p^2-1088p+8832)}{3267p^2} \ ,
\eeq
which is non-zero at spatial infinity.
The singularity exists only at $p=0$, hence
the solution \eqref{metric5d_0_mag} is regular at $p=2$.
Therefore, the solution obtained by the uplifting is a black hole.

The horizon of this black hole is known as the Nil manifold
in the Bianchi classification~\cite{Kodama:1997tk}
and has a negative constant Ricci scalar.
With this result at hand, it is natural to expect that other
Bianchi type horizon may be obtained in a similar fashion.
In fact, we will find the Bianchi type IV type horizon
in the next subsection. It will be interesting to examine
if all of the possible 3-dimensional geometry
in the Thurston's classification~\cite{Thurston:1982zz}
appear as the horizon geometries of 5-dimensional black holes.

\subsubsection{AdS black holes with a lens space topology}

Let us write the magnetic solutions corresponding to
the non-planar solution \eqref{CHMiii}.
If $\kappa=1$, the duality relation \eqref{dualpotential}
gives
\bea
 \partial_\theta \omega (\theta) = -e^\Omega \alpha^{-1} A' \sin \theta=-Q \sin \theta \ .
\eea
Hence, the dual magnetic potential is
\beq
	\mathcal{\omega} =  Q \textrm{cos} \theta \ud \varphi .
\eeq
Since the 4-dimensional metric and the dilaton \eqref{CHMiii}
do not change at all by the duality transformation,
the 5-dimensional metric is now given by
\bea
\ud s^2_{5D}&=&-e^{\frac{\Phi_0}{\sqrt3}}\Big(\ba_Nx^2+ x-\frac83M\Big)\frac{\ud t^2}{x}
+\frac{e^{\frac{\Phi_0}{\sqrt3}}\,\ud x^2}{\ba_Nx^2+ x-\frac83M} \nonumber \\
	&&+e^{\frac{\Phi_0}{\sqrt3}}x
      \left[ \Big(\ud \theta^2+ \sin^2 \theta \ud\varphi^2 \Big)
    +e^{-\sqrt{3} \Phi_0}
    \Big(\ud w + Q \textrm{cos} \theta \ud\varphi\Big)^2 \right].
    \label{lens-BH}
\eea
With a change of variable $r=\sqrt{x}$, it is easy to see the solutions
are AdS black holes.
Actually, the black hole \eqref{lens-BH} is an AdS black hole with a lens space topology.
Using the invariant basis
\begin{equation}
 \begin{split}
  \sigma^1 &= -\sin w d\theta + \cos w \sin\theta d\varphi\ , \\
  \sigma^2 &= \cos w d\theta + \sin w \sin\theta d\varphi\ , \\
  \sigma^3 &= d w + \cos\theta d\varphi \ ,
 \end{split}
\end{equation}
we can represent the metric of a 3-sphere as
$$
(\sigma^1)^2 + (\sigma^2)^2 + (\sigma^3)^2
= d\theta^2 + \sin \theta^2 d\varphi^2 + (\ud w + \cos \theta \ud \varphi )^2 \ .
$$
The above parametrization appears when we discuss the Bianchi type IX spacetime.
The horizon geometry in the above black hole solution \eqref{lens-BH} looks like
$$
\ud\theta^2 + \sin \theta^2 \ud\varphi^2
+ e^{-\sqrt{3} \Phi_0}(\ud w + Q \cos \theta \ud\varphi )^2 \ .
$$
By rescaling the coordinates and the integration constants, we can set $\Phi_0 =0$.
The charge $Q$ has to be quantized to an integer from regularity requirements.
This quotient space is the so-called lens space. Thus, we have a black hole
with a non-trivial topology by using our solution-generating method.

\subsubsection{Squashed Kaluza-Klein black holes}

In the $\Lambda =0$ case, we have more interesting solutions.
Here, the horizon is Bianchi type IV again. However, this time, the squashing parameter
depends on the radial coordinate. Hence, it can not be absorbed
 by a simple rescaling of coordinates. The horizon of the black hole is genuinely squashed. 
First of all, we need to write down the 4-dimensional solution for $\Lambda =0$.
In the previous section, we skipped this simple exercise.
Starting from the ansatz for the vector potential
\bea
  A = \frac{A_0}{x} \ ,
\eea
we can solve our equations of motion for the case $\gamma\delta =-1$.
Indeed, it is easy to derive the 4-dimensional metric
\bea
 ds^2 =
  -\frac{x-x^-}{\sqrt{x} \sqrt{x + x^+}} dt^2
  + \frac{ \sqrt{x} \sqrt{x-x^+} }{x + x^-} dx^2
  + x^{3/2} \sqrt{x+x^+} \left( d\theta^2 + \sin^2 \theta d\varphi^2 \right)
\eea
and the dilaton
\bea
 e^{\delta \Phi} = \sqrt{\frac{x}{x+x^+}}  \ ,
\eea
where $x^\pm$ are  constants of integration.
Next, we have to obtain dual magnetic solutions with $\gamma \delta =1$.
Again, the metric and the dilaton do not change by the dual transformation.
And, irrespective of the ansatz for $A$,
the dual magnetic potential is given by
\beq
	\mathcal{\omega} =  Q \textrm{cos} \theta \ud \varphi .
\eeq

Uplifting the above solutions, we find the 5-dimensional metric
\bea
	\ud s^2_{5D} &=&
      -\left( 1- \frac{x^-}{x} \right) \;\ud t^2
      + (1+\frac{x^+}{x}) \left(1-\frac{x^-}{x}\right)^{-1} \ud x^2
      +x^2(1+\frac{x^+}{x})\l[\ud \theta^2 + \sin^2 \theta \ud\varphi^2\r] \nonumber \\
	&& +\left(1+\frac{x^+}{x}\right)^{-1} \l(\ud w + Q \cos \theta \ud\varphi\r)^2.
\eea
Notice that this solution locally looks like a black string
when $x^+ =0$.

The solutions we have obtained
are nothing but squashed Kaluza-Klein black holes which look like
5-dimensional black holes in the vicinity of the horizon, but 4-dimensional
Minkowski spacetime times a circle in the far region~\cite{squash}.
It has been shown that these solutions are stable~\cite{Kimura:2007cr}.
Since the Hawking radiation carries the information of the horizon,
squashed Kaluza-Klein black holes could be a window to
 extra-dimensions~\cite{Ishihara:2007ni}.

So far, we have discussed 5-dimensional black holes
obtained by uplifting 4-dimensional black holes.
It is easy to extend the above analysis to higher dimensions.
There, we might find more interesting solutions by using our general
solution generating method.

\section{Conclusions}

In this paper, we have studied the system of equations of motion derived from Einstein-Maxwell-Dilaton theories with a Liouville potential. Although this system is fully integrable when the cosmological constant is set to zero \cite{gibbonsmaeda} (no potential), this is no longer true when the potential coupling $\Lambda$ is switched on. We wrote the $d$-dimensional system of equations of motion for two particular symmetries : cylindrical ($\ka=0$) and $d-2$ maximally symmetric subspaces ($\ka\neq0$). In both cases, we pushed the integration of the system as far as possible reducing the number of equations or unknown variables. For the cylindrical metrics, we reduced the problem to solving a unique non-linear second-order differential equation. Once a solution to this equation is found,  the whole metric solution can be derived and constrained to be a black hole solution of the modified Einstein equations by fixing some of the integration constants. It is worth noting that for specific values of the couplings, $\ga\da=1$, we found that the system is \emph{fully} integrable. This case contains also the specific subclass of solutions which can be uplifted to be full 5-dimensional solutions to the 5-dimensional Einstein theory. In the general case, where the general solution cannot be found (at least in the framework developed here), we explored polynomial solutions, keeping in mind that we aimed for black hole spacetimes.

The generic solutions to the theory are inhomogeneous metrics (that is, not black holes) with several (two to three) naked singularities. However, black holes are contained as a subclass, once two of the integration constants are interrelated. Then, we get usually \emph{one event horizon}, except in the string case where there are \emph{two}. The horizons have planar topology and the solutions are neither asymptotically dS or adS, except within the case $\da=0$, as expected from \cite{wiltshire}. However, this comes at the cost of having a constant dilaton.

The situation is quite different in the case of maximal symmetry. First of all, black hole solutions are readily obtained without need for further tampering with the integration constants, which are fixed accordingly to the cylindrical case by a previously absent off-diagonal equation of motion. Second of all, the system is never fully integrable and can only be reduced to two second-order non-linear coupled differential equations. In the special case $\ga\da=1$, these equations decouple and an exact expression for the Maxwell field can be obtained. Nevertheless, there always remains at least one non-integrable equation, for which ans\"atze have to be provided. Then, polynomial solutions can be found with integer or non-integer exponents, yielding one or two horizons depending on cases. A criterium for non-zero cosmological constant in this system of coordinates was also derived. The solutions, in agreement with \cite{CHM}, all have unusual asymptotics.

\begin{table}[h]
\begin{tabular}{|c|c|c|c|c|c|}
	\hline
	& Values of constants & $g_{tt}$ & $g_{qq}$ & $e^{\Phi}$&Solutions found\\
	\hline
	$K_{1,2}$ & $\sig=1$, $\ka>0$ & $q^{\frac2{\ga^2}}$ & $ct$ & $q^{\pm\frac2\ga}$& \eqref{CHMii}\\
	\hline
	$M_{1,2}$ & $\sig\ka>0$ & ct & ct & ct &\\
	\hline 
	$N_{1,2}$ & $\da^2<3$, $\sig\Lambda<0$, $\ka=0$ & $q^2$ & $q^{2(\da^2-1)}$ & $q^{2\da}$&\eqref{solution:2}, \eqref{solution:4}, \eqref{solution:5},\eqref{solution:9}, \eqref{solution:10}, \eqref{CHMiii} \\
	\hline
	$P_{1,2}$ & $\sig\Lambda<0$, $sign(\ka)=sign\big(\sig(\da^2-1)\big)$ & $q^{\frac2{\da^2}}$ & ct & $q^{\frac2\da}$ &\eqref{CHMi}\\
	\hline
	$T_{1,2}$ & $\sig\Lambda w<0$, $\ka=0$, $\sig uv<0$ & $q^{2\frac{\ga^2-\da^2+4}{(\ga-\da)^2}}$ & $q^{2\frac{\da+\ga}{\da-\ga}}$ & $q^{\frac4{\da-\ga}}$&\eqref{solution:3}, \eqref{solution:7} \\
	\hline
\end{tabular}
\caption{Asymptotic form of solutions for trajectories approaching critical points at phase space infinity from within the sphere at infinity, in the case $q\to\infty$}
\label{TableWilt}
\end{table}

Table \ref{TableWilt} is taken from \cite{wiltshire} and classifies the various solutions by their global asymptotic properties, when put under the form :
\beq
	\ud s^2 \sim -g_{tt}\ud t^2 + g_{qq}\ud q^2 +q^2\l(\ud \theta^2+\si(\theta)^2\ud\varphi^2\r).
\eeq
Thus, each category $K_{1,2}$, $M_{1,2}$, $N_{1,2}$, $P_{1,2}$, $T_{1,2}$ corresponds to points in the phase space of the solutions, attracting or repulsing solution trajectories. This way, the solutions are classified according to their asymptotical behaviour, see Table \ref{TableWilt} (for a much more detailled analysis, see \cite{wiltshire}).

We have filled in that same table with the solutions we obtained. No solution corresponds to the points $M$ since we took special care to avoid flat space. We see a clear separation in the cylindrical $\ka=0$ solutions, since all - general $\ga\da=1$ or polynomial - solutions with non-linear Maxwell fields belong to the class $N$, are asymptotically adS iff $\da=0$, and as has been discussed at length, depend crucially in their behaviour on the sign of $\da^2-3$; whereas other solutions, with linear Maxwell field - polynomial or with $\eta=0$ - belong to the class $T$, which can also be regular asymptotically in the limit $\da=0$ and nowhere else. The class $N$ has an equivalent if $\da^2>3$ (which we do not reproduce here for the sake of brevity), reproducing the separation of cases of section \ref{subsubsection:ka_0_generalBH}.

The $\ka\neq0$ solutions are classified in the same manner : they can belong to different classes and even have endpoints in $\ka=0$ subclasses [\eqref{CHMiii}]. Globally, there is a good match between global properties as determined by \cite{wiltshire} and the exact solutions we listed.

The system of equations of motion was solved explicitly for electric Maxwell fields without any loss of generality. Indeed, a duality procedure \cite{CLSZ}, was generalized and used to find magnetic equivalents of all the electric solutions derived. Then, various 5-dimensional electric metrics and dualized magnetic metrics were written and analyzed.

Several open questions remain : though we have not found explicit  solutions for regular asymptotic dS or adS solutions, their existence was hinted at both pertubatively and numerically (\cite{wiltshire}) in the pure cosmological constant case ($\delta=0$). For $\da\neq0$, a no-go theorem was formulated. However, we could not obtain non-trivial solutions for $\da=0$, nor the equivalent of a no-go theorem for non-constant dilatonic solutions. This is not necessarily in contradiction with \cite{wiltshire}, since only the asymptotic behaviour of solutions is predicted, and thus it is consistent with our result that, in order to get an asymptotically adS solution, the scalar field has to be frozen everywhere, thereby restoring the regularity of the asymptotics. Furthermore, no $\kappa\neq 0$ topology black hole solution could be obtained for the string case in the electric case  though there does not seem to be any fundamental argument against its existence.

Perspectives and extensions of our work include the generalization of the solutions to D dimensions by making full use of the D-dimensional equations of motion of section 2. Such solutions might be relevant for supergravity setups in higher dimensions for instance. The inclusion of a Gauss-Bonnet term in the 5-dimensional action could also be considered, as Lovelock theory (for a review see \cite{Charmousis}) is the natural generalisation of Einstein theory in higher dimensions and may well regularise the asymptotics. Perturbative and numerical results (\cite{ohta}) were carried out but no analytical results were presented up to now. It has already been hinted upon that GR-like behaviour may result from a scalar-tensor theory {\it given} the addition of higher order terms dictated by Kaluza-Klein reduction of Lovelock theories \cite{Amendola}.

\acknowledgements{
We wish to thank Paul-Thomas Desessarts for collaboration  at the early stages of this project. CC wishes to thank the Department of Physics in Kyoto University and the Yukawa Institute for Theoretical Physics for hospitality during the completion of this work.
JS is supported by the Japan-U.K. Research Cooperative Program,
Grant-in-Aid for  Scientific Research Fund of the Ministry of
Education, Science and Culture of Japan No.18540262.
}

\end{document}